\documentclass[aps,prd,preprint,nofootinbib,eqsecnum,12pt]{revtex4}
\usepackage{amsfonts,amsmath,amssymb,graphicx,bbm,latexsym,slashed}

\def\beq{\begin{equation}}
\def\eeq{\end{equation}}
\newcommand{\beqa}{\begin{eqnarray}}
\newcommand{\eeqa}{\end{eqnarray}}
\newcommand{\barr}{\begin{array}}
\newcommand{\earr}{\end{array}}
\def\nn{\nonumber}

\newcommand{\Glf}{{\rm G}_{\rm {LF}}}
\newcommand{\su}{{\rm {SU}}}
\newcommand{\tr}{{\rm {Tr}}}

\begin{document}

\rightline{\today}

\title{Minimal Lepton Flavor Violation and Renormalization Group Evolution of Lepton Masses and Mixing 
}

\def\cornell{Institute for High Energy Phenomenology\\
Newman Laboratory of Elementary Particle Physics\\ Cornell University,
Ithaca, NY 14853, USA}

\author{Yuval Grossman}
\email{yg73@cornell.edu}
\affiliation{\vspace*{4mm}\cornell\vspace*{6mm}}

\author{Shamayita Ray}
\email{sr643@cornell.edu}
\affiliation{\vspace*{4mm}\cornell\vspace*{6mm}}

%\pacs{11.10.Hi, 11.30.Hv, 14.60.Pq, 14.60.St}

%%%%%%%%%%%%%%%%%%%%%%%%%%%%%%%%%%%%%%%%%%%%%%%%%%%%%%%%%
\begin{abstract}

\vspace*{4mm}

We study the renormalization group equations (RGEs) of the neutrino
parameters in models of Minimal Lepton Flavor Violation. In such
models, the RGEs can be described in terms of flavor spurions, such
that only the coefficients depend on the specific model. We explicitly
demonstrate this method for the SM and MSSM for both Type-I and 
Type-III seesaw models. For that purpose, the RGEs of neutrino 
parameters in the MSSM Type-III seesaw have been computed. 
We have extended this method to get the evolution equations at 
second order. The implications for leptogenesis are also discussed.

\end{abstract}

\pacs{}

\keywords{}

\preprint{}

\maketitle

%%%%%%%%%%%%%%%%%%%%%%%%%%%%%%%%%%%%%%%%%%%%%%%%%%%%%%%%%
%%%%%%%%%%%%%%%%%%%%%%%%%%%%%%%%%%%%%%%%%%%%%%%%%%%%%%%%%
\section{Introduction}
\label{intro}
%%%%%%%%%%%%%%%%%%%%%%%%%%%%%%%%%%%%%%%%%%%%%%%%%%%%%%%%%
%%%%%%%%%%%%%%%%%%%%%%%%%%%%%%%%%%%%%%%%%%%%%%%%%%%%%%%%%

The data from past and ongoing neutrino oscillation experiments, as
well as from cosmology and astrophysics, have now confirmed that
neutrinos have distinct masses and that the three neutrino flavors
$\nu_e$, $\nu_\mu$ and $\nu_\tau$ mix among themselves to form
the three mass eigenstates. The fact that the neutrinos are massive
and mix implies non-conservation of lepton flavor. Hence, lepton
flavor violating processes are expected in the lepton sector
just as quark flavor violating processes arise in the quark sector.

In the quark sector of the Standard Model (SM), flavor violation is
induced by the Yukawa matrices such that baryon number remains an
exact symmetry. The fact that flavor changing neutral currents (FCNCs)
are heavily suppressed puts stringent constraints on the possible
structure of new degrees of freedom carrying flavor quantum
numbers. These constraints can be satisfied either if
new particles are very heavy or if flavor symmetries suppress the
flavor changing couplings. One of the most predictive and restrictive
symmetry principles that can be used is Minimal Flavor Violation
(MFV)~\cite{MFV}. The MFV framework is the assumption that
in the quark sector the only sources of flavor symmetry breaking are
the Yukawa couplings. 

While the idea of MFV has a straightforward and unique realization in
the quark sector, the situation is different in the lepton sector. The
reason is that the neutrinos can be Majorana particles, in which case total
lepton number is no longer a symmetry of the theory. Due to this
complication, the Minimal Lepton Flavor Violation (MLFV) hypothesis is
not uniquely defined; there are two ways to define it~\cite{MLFV}. In
the first case, known as MLFV with minimal field content, we do not
add any new field to the theory, and treat the neutrino mass terms as
non-renormalizable terms. The only irreducible sources of lepton
flavor violation are the charged-lepton Yukawa matrix and the
effective left-handed Majorana mass matrix. The breaking of total
Lepton Number (LN) is independent of the flavor violation and happens
at some very high scale. 

The other possibility, called MLFV with extended field contents
(MLFV-ex), is to introduce new fields to the SM. In particular, three
heavy right-handed neutrinos are added to the SM. Their
Majorana mass term, which is assumed to be flavor universal,
explicitly breaks LN. In this scenario, the two Yukawa matrices act as
the only irreducible sources of flavor violation. In the MLFV-ex
scenario, the low energy observables depend on the high energy
parameters of the theory. For example, the FCNC constraints in the
leptonic sector affects leptogenesis. This has been studied in
\cite{MLFV-leptogenesis-Isidori} with the mass-splitting of the
right-handed neutrinos, required for successful leptogenesis, being
introduced from flavor symmetry considerations only. To have a
complete understanding of the relation between the high energy
parameters and the low energy observables, one needs to study the
complete renormalization group (RG) evolution effects in this context.
RG evolution has already been shown to have strong effects on
leptogenesis \cite{radiative-leptogenesis}.
Ref.~\cite{MSSM-MLFV-RG} shows the
stability of the MLFV under RG evolution in the context of soft masses
in the Minimal Supersymmetric Standard Model (MSSM). While
\cite{MLFV-pheno} takes into account the RG evolution effects in the
context of $\mu \to 3e$ and $\tau \to 3\ell$ decays, a general
analysis of RG evolution of lepton masses and mixing parameters in the
MLFV framework is still lacking. 

In this paper, we consider the RG evolution of lepton masses and 
mixing parameters in the MLFV-ex scenario, with the SM as the low 
energy effective theory. The basic idea is that the RGEs can be 
written in terms of spurions that depend only on the Yukawa matrices. 
The coefficient of each term can be model dependent. 
Moreover, we assume that the universality of the Majorana masses is
broken slightly, and hence treat the Majorana mass matrix as a spurion
of our theory, and we get the RGE for this spurion as well. This is,
in fact, a natural assumption as the universality is automatically
broken in course of RG evolution. We show explicitly how one can 
write the RGEs for the SM and MSSM in both Type-I and Type-III seesaw
models. The advantage of the spurion formalism is that it shows how
each combination enters and can be used as a check for any MLFV model.

%%%%%%%%%%%%%%%%%%%%%%%%%%%%%%%%%%%%%%%%%%%%%%%%%%%%%%%%%
%%%%%%%%%%%%%%%%%%%%%%%%%%%%%%%%%%%%%%%%%%%%%%%%%%%%%%%%%
\section{The Model: $\nu$MSM and MLFV with extended field content}
\label{sec:model}
%%%%%%%%%%%%%%%%%%%%%%%%%%%%%%%%%%%%%%%%%%%%%%%%%%%%%%%%%
%%%%%%%%%%%%%%%%%%%%%%%%%%%%%%%%%%%%%%%%%%%%%%%%%%%%%%%%%

We consider the SM extended by three right-handed neutrinos, which are
singlets under the SM gauge group. 
This model is referred to as the $\nu$MSM~\cite{nuMSM}. 
We also consider the case where they are
triplet under the SU(2)$_L$ group later in this section.

We begin by considering the model excluding all mass 
terms of the leptons and gradually introduce mass terms to 
study their effect on the flavor symmetries of the theory, 
at different energy scales $\mu$. 
In the massless lepton limit, at high scale $\mu > M_R$, 
the $\nu$MSM enjoys a flavor
symmetry $\Glf^0$, similar to that of the quark sector, given by
\beq
\Glf^0 = \su(3)_{l_L} \otimes \su(3)_{e_R} \otimes \su(3)_{\nu_R} \; .
\eeq 
Here we consider only the non-Abelian part of the flavor symmetry 
group. This sector is also invariant under U(1) of hypercharge ($Y$), 
total lepton number (LN), as well as U(1)$_{\rm E}$ (or U(1)$_{\nu}$), which 
corresponds to a rotation of the $e_R$ (or $\nu_R$) fields. 

The presence of Majorana mass term for the right-handed neutrinos
reduces the symmetry. Let us denote the right-handed neutrinos by
$\nu_R^i$, $i \in \{ 1,2,3\}$. The only source
of LN violation in this model is the Majorana mass term of these
right-handed neutrinos given by
\beqa
{\cal L}_{\rm Maj} = -\frac{1}{2} \bar\nu_R^{C} M_\nu \nu_R 
+ {\rm {h.c.}} \; , 
\label{L-Maj}
\eeqa
where $C$ denotes charge conjugation. The right-handed Majorana mass
matrix $M_\nu$ is symmetric, $M_\nu = M_\nu^T$. Furthermore, without
any loss of generality, we can choose $M_\nu$ to be real by
re-definition of the phases of $\nu_R^i$. (The $\nu$MSM was originally
defined \cite{nuMSM} in the basis where the charged lepton mass matrix
and the Majorana mass matrix are real and diagonal.) In
general $M_\nu$ breaks $\su(3)_{\nu_R}$ completely. For a
universal mass matrix, however, the breaking is into an O(3) group. In this
case, the Majorana mass matrix is given by
\beq
\left( M_\nu \right)_{ij} = M_R \delta_{ij} \; ,
\label{Mnu-uni}
\eeq
and the flavor symmetry group becomes 
\beq
\Glf^0 \to \Glf = \su(3)_{l_L} \otimes \su(3)_{e_R} \otimes {\rm O}(3)_{\nu_R} \; .
\eeq
The two Yukawas $Y_e$ and $Y_\nu$ are given by
\beqa
{\cal L}_{\rm Yukawa} = -  \bar e_R Y_e \phi^\dagger l_L 
-  \bar \nu_R Y_\nu {\widetilde\phi}^\dagger l_L + {\text {h.c.}} \; ,
\label{L-Yuk}
\eeqa
where $\phi$ is the SM Higgs doublet and $\widetilde\phi = i \sigma^2 \phi^\ast$, 
$\sigma^2$ being the second Pauli matrix.

It is customary to treat $\Glf$ as an unbroken symmetry of the
underlying theory which can be achieved by treating the Yukawa 
matrices as spurion fields with non-trivial quantum
numbers under $\Glf$
\beq
Y_e \sim (\bar 3, 3, 1) \; ,  \quad \quad Y_\nu \sim (\bar 3, 1, 3) \; . 
\eeq
The $\nu$MSM in the massless lepton limit and with 
universal right-handed Majorana masses enjoys the flavor symmetry 
$\Glf$ and this is the MLFV hypothesis with extended 
field content (MLFV-ex)~\cite{MLFV}.
Going beyond the MLFV-ex hypothesis, in this paper we choose the
universality of $M_\nu$ to be slightly broken, which happens also as a
result of RG evolution. We thus treat $M_\nu$ as a spurion transforming, 
under $\Glf$, as
\beq
M_\nu \sim (1, 1, 6)  \; .
\eeq
The spurions have the following transformation properties:
\beqa
Y_e \to U_R Y_e U_L^\dagger \; , \qquad Y_\nu \to O_\nu Y_\nu
U_L^\dagger \; , \qquad
M_\nu \to O_\nu M_\nu O_\nu^T \; ,
\label{Ye-Ynu-transform}
\eeqa
where $U_L \in \su(3)_{l_L}$, $U_R \in \su(3)_{e_R}$ and $O_\nu \in
{\rm O}(3)_{\nu_R}$. This technique is known as spurion analysis.

Finally, the heavy fields generate small neutrino masses via the
seesaw relation~\cite{type-I-seesaw}
\beq
m_\nu = \frac{v^2}{2} Y_\nu^T M_\nu^{-1} Y_\nu 
\; ,\label{mnu-Glf}
\eeq
where the vacuum expectation value of the SM Higgs is defined as
$\langle \phi \rangle = (0, v/\sqrt{2} )^T$. In the MLFV-ex model, 
the left-handed neutrino mass matrix is given by 
\beq
\left. m_\nu \right\arrowvert_{\rm MLFV-ex} = \frac{v^2}{2 M_R} Y_\nu^T Y_\nu \; .
\eeq 
Note that in general $Y_\nu^T Y_\nu$ and $Y_\nu^\dagger Y_\nu$ are two 
different sources of $\Glf$ breaking. Only in the limit where 
$Y_\nu$ is real are they the same \cite{MLFV}. We do, however, 
expect to have CP violation in the theory and thus we do not 
concentrate on the case of real $Y_\nu$. We consider the MLFV-ex model 
for $\mu > M_R$ energy regime for the rest of the paper, 
with the exception that the universality of $M_\nu$ is assumed 
to be slightly broken. We consider the case where $M_R$ is large 
compared to the electroweak symmetry breaking scale. This ensures 
that U(1)$_{\rm {LN}}$ is broken at some high scale, and that, 
in general, the breaking of LN by the Majorana mass term is 
independent of $\Glf$-violation.

Next, we discuss the effective theory below $M_R$, or 
equivalently below the scale of the lightest of the heavy 
right-handed neutrinos, when universality is broken. 
In this regime, all the three heavy right-handed 
neutrinos get integrated out, and as a result the flavor symmetry group 
reduces to
\beq
\Glf \to \Glf^\prime = \su(3)_{l_L} \otimes \su(3)_{e_R} \; .
\eeq
In this energy region, the dimension-5 non-renormalizable term in the
Lagrangian responsible for the LN-violating left-handed Majorana
neutrino masses is of the form 
\beq
{\cal L} \sim \bar l_L^C m_\nu l_L \phi \phi \; .
\eeq 
There are two sources of $\Glf^\prime$ breaking in this case. 
The charged lepton Yukawa $Y_e$ and the left-handed neutrino mass
$m_\nu$ that transform as
\beq
Y_e \sim (\bar 3,3)\;,\quad m_\nu \sim (6, 1)\;.
\eeq 
Thus, the model becomes equivalent to the MLFV hypothesis with minimal
field content~\cite{MLFV}. In this case, $m_\nu$ remains the only
relevant quantity that contains the high energy information of the
neutrino parameters, which in turn can be extracted by the measurement
of the neutrino masses and mixing parameters. Hence, the effect of RG 
evolution becomes an important factor to be taken into account, which 
we will be studying in the following sections.

In the framework of spurion analysis, $\Glf$ is broken by the
background values of the spurions. We consider the background values 
of $Y_{e,\nu}$ to be small, the largest one being experimentally 
measured to be $Y_\tau \sim 0.01$ at the scale $M_Z$. Thus, we can 
use perturbation theory and consider only the leading order 
corrections. To first order, the operators responsible for the 
breaking of $\Glf$ are combinations of two Yukawa matrices, 
that is, working at one loop is equivalent of considering 
spurions with two couplings. There are several combinations 
of couplings that can appear in the result. These
couplings and their transformation properties are given in
Table~\ref{tab:trans}. As can be seen, $M_\nu$ appears only 
when we consider the evolution of $M_\nu$ itself.
\begin{table}[t!]
\begin{center}
\begin{tabular}{|c|c|}
\hline
Combination of spurions & Transformation \\
\hline
$Y_e^\dagger Y_e$ & $(8 \oplus 1, 1, 1)$ \\
\hline
$Y_e Y_e^\dagger $ & $(1, 8 \oplus 1, 1)$ \\
\hline
$Y_\nu^\dagger Y_\nu$ & $(8 \oplus 1, 1, 1)$ \\
\hline
$Y_\nu Y_\nu^\dagger$ & $(1, 1,  8 \oplus 1)$ \\
\hline 
$\tr [Y_e^\dagger Y_e] = \tr [Y_e Y_e^\dagger]$ & $(1, 1, 1)$ \\
\hline 
$\tr [Y_\nu^\dagger Y_\nu] = \tr [Y_\nu Y_\nu^\dagger]$ & $(1, 1, 1)$ \\
\hline
$T_e \equiv Y_e^\dagger Y_e - \frac{1}{3} \tr [Y_e^\dagger Y_e] {\mathbf I}_3$ &  $(8 , 1, 1)$ \\ 
\hline
$T_e^\prime \equiv Y_e Y_e^\dagger - \frac{1}{3} \tr [Y_e Y_e^\dagger] {\mathbf I}_3$ &  $(1,8,1)$ \\
\hline
$T_\nu \equiv Y_\nu^\dagger Y_\nu - \frac{1}{3} \tr [Y_\nu^\dagger Y_\nu] {\mathbf I}_3$ &  $(8 , 1, 1)$ \\ 
\hline
$T_\nu^\prime \equiv Y_\nu Y_\nu^\dagger - \frac{1}{3} \tr [Y_\nu Y_\nu^\dagger] {\mathbf I}_3$ &  $(1, 1, 8)$ \\ 
\hline
\end{tabular}
\end{center}
\caption{Transformations of combinations of two spurion 
fields under $\Glf$. We have used the $\su(3)$ algebra $3 \otimes \bar 3 = 8 \oplus 1$.
\label{tab:trans}}
\end{table}

The flavor symmetry structure is more complicated when the heavy
neutrinos are not exactly degenerate. A breaking of the universality of
$M_\nu$, however small, is also necessary for leptogenesis as has been
shown in~\cite{MLFV-leptogenesis-Isidori}. In that paper, the degeneracy is
broken by appropriate combinations of spurions in the MLFV-ex
scenario. Our assumption is that the amount of non-degeneracy is
small and $\Glf$ is still the flavor symmetry of the underlying theory. 
The effect of the breaking is due to the fact that running at
the scale in between the three masses is not described by any of the
two regions we discussed above. Yet, if the breaking is small this 
running is not significant and integrating out all the neutrinos
together is a good approximation. Moreover, if the degeneracy is
lifted due to RG evolution, then taking it into account is formally
a higher order effect.

%%%%%%%%%%%%%%%%%%%%%%%%%%%%%%%%%%%%%%%%%%%%%%%%%%%%%%%%%
%%%%%%%%%%%%%%%%%%%%%%%%%%%%%%%%%%%%%%%%%%%%%%%%%%%%%%%%%
\section{RG evolution of neutrino parameters}
\label{sec:RG}
%%%%%%%%%%%%%%%%%%%%%%%%%%%%%%%%%%%%%%%%%%%%%%%%%%%%%%%%%
%%%%%%%%%%%%%%%%%%%%%%%%%%%%%%%%%%%%%%%%%%%%%%%%%%%%%%%%%

We now study the effect of RG running. At energy scales
above $M_R$, the quantities of interest are the Yukawa matrices $Y_{e,
\nu}$, and the right-handed neutrino mass matrix $M_\nu$. Below, we see
how they run.

In all our discussions, we consider only one loop running. In term of
spurions, each loop add two Yukawa terms, and thus working at one loop
is done by using only terms that have two Yukawa couplings more than
the tree level one. The evolution equations at second order are 
discussed in Appendix~\ref{app:2loop}.

%%%%%%%%%%%%%%%%%%
\subsection{RG evolution of $Y_{e}$}
\label{sec:RG-Ye-1}

We define
\beq
\dot Y_e \equiv \frac{d Y_e}{dt} \; , \qquad
 t \equiv \frac{\ln{(\mu/\mu_0)}}{16 \pi^2} \; .
\eeq 
Here $\mu_0 (> M_R)$ is some high energy scale at which we start running 
and the factor $(16\pi^2)$ appears because of the fact that we 
consider radiative corrections at 1-loop.

Under the flavor symmetry group $\Glf$, $Y_e$ transforms as $( \bar
3,3,1)$ and so does $\dot Y_e$. Hence $\dot Y_e$ can be expressed as
appropriate combinations of the spurion fields transforming as $(\bar
3,3,1)$.
Table~\ref{tab:trans} shows the combinations of two spurion fields with 
their transformation properties. Using the $\su(3)$ algebra 
\beq
8 \otimes \bar 3 = \overline{15} \oplus 6 \oplus \bar 3 \; , \qquad
8 \otimes 3 = 15 \oplus \bar 6 \oplus 3 \; ,
\label{su3-algebra}
\eeq
and we can write
\beqa
Y_e T_e &=& (\bar 3, 3, 1) \otimes (8,1,1) \ni (\bar 3, 3, 1) \; , \\
Y_e T_\nu &=& (\bar 3, 3, 1) \otimes (8,1,1) \ni (\bar 3, 3, 1) \; , \\
Y_e \tr [Y_e^\dagger Y_e] &=& (\bar 3, 3, 1) \otimes (1,1,1) = (\bar 3, 3, 1) \; , \\
Y_e \tr [Y_\nu^\dagger Y_\nu] &=& (\bar 3, 3, 1) \otimes (1,1,1) = (\bar 3, 3, 1) \; . 
\eeqa
The above combinations are the only terms, containing three
spurion fields, allowed to appear on the right-hand side (RHS) of the 
RGE for $\dot Y_e$. $T_e^\prime Y_e$ gives the same term as that given by $Y_e T_e$
and so has not been listed separately. Thus, at 1-loop, when terms up
to combinations of three spurion fields are allowed, the most general
form of $\dot Y_e$ is given by
\beqa
\dot Y_e &=& \widetilde{a_1} Y_e T_e + \widetilde{a_2} Y_e T_\nu
+ \widetilde{a_3} Y_e \tr[Y_e^\dagger Y_e] 
+ \widetilde{a_4} Y_e \tr[Y_\nu^\dagger Y_\nu] 
+ a_5 Y_e 
\label{Ye-1a} \nn \\
&=& Y_e \left( a_1 Y_e^\dagger Y_e + a_2 Y_\nu^\dagger Y_\nu 
+ a_3 \tr[Y_e^\dagger Y_e] + a_4 \tr[Y_\nu^\dagger Y_\nu] + a_5 {\mathbbm 1}_3 \right) \; ,
\label{Ye-1b}  
\eeqa
where $a_1$, $a_2$, $a_3$, $a_4$ and $a_5$ are expected to be numbers of
${\cal O}(1)$ that can be determined by the calculation of the 1-loop
diagrams in the theory.

\begin{figure}[t!]
\begin{center}
\includegraphics[scale=1]{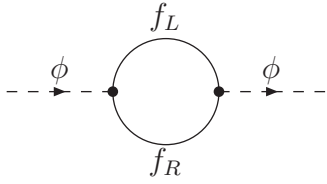}
\caption{The self-energy diagram of the Higgs $\phi$ with complete 
fermion loop, where the fermion pair $\{ f_L, f_R \}$ can be
$\{ l_L, e_R \}$, $\{ l_L, \nu_R \}$, $\{ q_L, u_R \}$ or
$\{ q_L, d_R \}$ producing contributions proportional to 
$\tr[Y_e^\dagger Y_e]$, $\tr[Y_\nu^\dagger Y_\nu]$, $\tr[Y_U^\dagger Y_U]$ 
and $\tr[Y_D^\dagger Y_D]$ respectively.
\label{Higgs-self}}
\end{center}
\end{figure}

The case of $a_5$ is a bit more involved since it is a function
independent of spurion fields. Thus $a_5$ must contain combinations of
other couplings in the theory that transform trivially under
$\Glf$. The couplings that we have in the theory are the gauge
couplings, $g_i$, the Higgs self-coupling, $\lambda$, and the quark
Yukawa couplings $Y_{U,D}$.
Since leptons are singlets under $\su(3)_C$, $g_3$ cannot
contribute. Moreover, at 1-loop the Higgs self-coupling cannot
contribute either. Terms proportional to $g_1$ and $g_2$ contributing to
$a_5$ must be of form
\beq
a_{g_1} g_1^2 + a_{g_2} g_2^2 \; .
\eeq
The singlet combination made of the quark Yukawas $Y_{U,D}$ 
is of the form $\tr[Y_i^\dagger Y_i]$, and the most 
general form of the quark Yukawa contributions to $a_5$ is
\beq
a_{U} \tr[Y_U^\dagger Y_U] + a_{D} \tr[Y_D^\dagger Y_D].
\eeq
Thus the general form of $a_5$ is given by
\beq
a_5 = a_{g_1} g_1^2 + a_{g_2} g_2^2 + a_{U} \tr[Y_U^\dagger Y_U] + a_{D} \tr[Y_D^\dagger Y_D] \; ,
\eeq
and the general form of $\dot Y_e$ becomes
\beqa
\dot Y_e &=& Y_e \left( a_1 Y_e^\dagger Y_e + a_2 Y_\nu^\dagger Y_\nu \right) 
+ Y_e \left( a_{g_1} g_1^2 + a_{g_2} g_2^2 \right) \nn \\
&+& Y_e \left( a_3 \tr[Y_e^\dagger Y_e] + a_4 \tr[Y_\nu^\dagger Y_\nu] + 
a_{U} \tr[Y_U^\dagger Y_U] + a_{D} \tr[Y_D^\dagger Y_D]\right) \; .
\label{Ye-gen-a4}
\eeqa
Terms proportional to $\tr[Y_x^\dagger Y_x]$ ($x \in \{e, \nu, U, D\}$) 
arise from a complete fermion loop 
in the self-energy correction of the scalar Higgs boson, as shown in 
Fig.~\ref{Higgs-self}. 
Since quarks come in three colors, one gets 
\beq
a_{3} : a_{4} : a_{U} : a_{D} = 1: r: 3: 3,
\label{trace-ratio}
\eeq
where each of the three quark colors contributes equally. $r \equiv
a_{4}/a_3$ is determined by the transformation properties of the
right-handed neutrinos under the gauge group. As we discuss below, at
Eq.~(\ref{r-SM}), for singlets $r = 1$, while for triplets $r=3$. We
can now define a quantity
\beqa
T \equiv \tr[Y_e^\dagger Y_e] + r \tr[Y_\nu^\dagger Y_\nu] + 3 \tr[Y_U^\dagger Y_U] + 3 \tr[Y_D^\dagger Y_D] \; ,
\label{T}
\eeqa
and write $\dot Y_e$ in a simpler form as
\beqa
\dot Y_e &=& Y_e \left( a_1 Y_e^\dagger Y_e + a_2 Y_\nu^\dagger Y_\nu \right) 
+Y_e \left( a_T T + a_{g_1} g_1^2 + a_{g_2} g_2^2 \right)   \; ,
\label{Ye}
\eeqa
where $a_T$, $a_{g_1}$, $a_{g_2}$, are expected to be of ${\cal O}(1)$.

%
%%%%%%%%%%%%%%%%%%%%%%%%%%%%%%%%%%%%%%%%%%%%%%%%%%%%%%%%%%%%%%%%%%%%
%%%%%%%%%%%%%%%%%%%%%%%%%%%%%%%%%%%%%%%%%%%%%%%%%%%%%%%%%%%%%%%%%%%%
\subsection{RG evolution of $Y_{\nu}$}
%%%%%%%%%%%%%%%%%%%%%%%%%%%%%%%%%%%%%%%%%%%%%%%%%%%%%%%%%%%%%%%%%%%%
%%%%%%%%%%%%%%%%%%%%%%%%%%%%%%%%%%%%%%%%%%%%%%%%%%%%%%%%%%%%%%%%%%%%

Next, we discuss the running of $Y_\nu$. Since $Y_\nu$
transforms as $(\bar 3, 1, 3)$ under $\Glf$, so must be $\dot Y_\nu$.
From Table~\ref{tab:trans} and using Eq.~(\ref{su3-algebra}) we obtain
that the only allowed combinations of spurion fields at one loop
order are
\beqa
Y_\nu T_e &=& (\bar 3, 1, 3) \otimes (8,1,1) \ni (\bar 3, 1, 3) \; , \\
Y_\nu T_\nu &=& (\bar 3, 1, 3) \otimes (8,1,1) \ni (\bar 3, 1, 3) \; , \\
Y_\nu \tr [Y_e^\dagger Y_e] &=& (\bar 3, 1, 3) \otimes (1,1,1) = (\bar 3, 1, 3) \;, \\
 Y_\nu \tr [Y_\nu^\dagger Y_\nu] &=& (\bar 3, 1, 3) \otimes (1,1,1) = (\bar 3, 1, 3)  .
\eeqa
$T_\nu^\prime Y_\nu$ gives the same term as that given by 
$Y_\nu T_\nu$ and so has not been written here.
Finally, as with $\dot Y_e$, we can write
\beqa
\dot Y_\nu &=& \widetilde{b_1} Y_\nu T_e + \widetilde{b_2} Y_\nu T_\nu
+ \widetilde{b_3} Y_\nu \tr[Y_e^\dagger Y_e] 
+ \widetilde{b_4} Y_\nu \tr[Y_\nu^\dagger Y_\nu] 
+ \widetilde{b_5} Y_\nu 
\label{Ynu-1a} \; ,
\eeqa
which can be simplified, using a similar approach to that of the previous section, to get
\beqa
\dot Y_\nu &=& Y_\nu \left( b_1 Y_e^\dagger Y_e + b_2 Y_\nu^\dagger Y_\nu \right) 
+Y_\nu \left( b_T T + b_{g_1} g_1^2 + b_{g_2} g_2^2 \right)   \; ,
\label{Ynu}
\eeqa
where $T$ is defined in Eq.~(\ref{T}), and $b_1$, $b_2$, $b_T$,
$b_{g_1}$, $b_{g_2}$, are expected to be of ${\cal O}(1)$.

%%%%%%%%%%%%%%%%%%%%%%%%%%%%%%%%%%%%%%%%%%%%%%%%%%%%%%%%%%%%%%%%%%%%%%%%%%%%%%%
%%%%%%%%%%%%%%%%%%%%%%%%%%%%%%%%%%%%%%%%%%%%%%%%%%%%%%%%%%%%%%%%%%%%%%%%%%%%%%%
\subsection{RG evolution of the heavy right-handed Majorana mass $M_\nu$}
\label{sec:RG-Mnu}
%%%%%%%%%%%%%%%%%%%%%%%%%%%%%%%%%%%%%%%%%%%%%%%%%%%%%%%%%%%%%%%%%%%%%%%%%%%%%%%
%%%%%%%%%%%%%%%%%%%%%%%%%%%%%%%%%%%%%%%%%%%%%%%%%%%%%%%%%%%%%%%%%%%%%%%%%%%%%%%

Once we know the evolution of the Yukawa matrices, we can 
discuss the running of the physical masses. We consider the 
right handed neutrino mass term
\beq
{\cal L}_{\rm Maj} = -\frac{1}{2} \left[ \bar\nu_R^C M_\nu \nu_R + 
\bar \nu_R M_\nu^\dagger \nu_R^C \right] \; .
\eeq
We first discuss the evolution of $M_\nu$ below and later consider
$ M_\nu^\dagger$.

As already stated, $M_\nu$ transforms as (1,1,6) under $\Glf$ and thus
is symmetric under O(3)$_{\nu_R}$. Hence while considering the RG
evolution of $M_\nu$, the RHS must contain terms which has the same
transformation properties under $\Glf$.
Using the transformation rules in Table~\ref{tab:trans} and the 
$\su(3)$ algebra
\beqa
6 \otimes 8 = 24 \oplus \overline{15} \oplus 6 \oplus \bar 3 \; ,
\label{su3-mnu}
\eeqa
the allowed terms are obtained to be
\beqa
M_\nu T_\nu^\prime &=& (1,1,6) \otimes (1,1,8) \ni (1,1,6) \; , \\
M_\nu \tr[Y_e^\dagger Y_e] &=& (1,1,6) \otimes (1,1,1) = (1,1,6) \; , \\
M_\nu \tr[Y_\nu^\dagger Y_\nu] &=&  (1,1,6) \otimes (1,1,1) = (1,1,6) \; .
\eeqa
The quark Yukawas, $Y_{U,D}$, are expected to have contributions of form 
$\tr[Y_i^\dagger Y_i]$ ($i \in \{ U,D\}$).
In general, there will also be terms containing $g_i^2$ and $\lambda$.

To get the final form of the $Y_\nu$ dependence of $\dot M_\nu$ we
have to take into account the fact that $M_\nu$ is symmetric.
Symmetrizing we obtain
\beqa
&& \frac{1}{2} \left[ {\left( M_\nu Y_\nu Y_\nu^\dagger \right)}^{\alpha \beta} 
+ {\left( M_\nu Y_\nu Y_\nu^\dagger \right)}^{\beta \alpha}\right] + 
\frac{1}{2} \left[ {\left( M_\nu \right)}^{\alpha \beta}\tr[Y_\nu^\dagger Y_\nu] 
+ {\left( M_\nu \right)}^{\beta \alpha}\tr[Y_\nu^\dagger Y_\nu] \right] \nn \\
&=& \frac{1}{2} \left[ {\bigl(M_\nu Y_\nu Y_\nu^\dagger \bigr)}^{\alpha \beta} 
+ {\bigl( {\left(Y_\nu Y_\nu^\dagger \right)}^T M_\nu \bigr)}^{\alpha \beta}\right] 
+ {\left( M_\nu \right)}^{\alpha \beta}\tr[Y_\nu^\dagger Y_\nu] \nn \; ,
\eeqa
where $\alpha, \beta$ are O(3)$_{\nu_R}$ indices.
We can then write the most general form of the RG equation for 
$M_\nu$ as
\beqa
\dot M_\nu &=& \frac{q_1}{2} \left[ M_\nu (Y_\nu Y_\nu^\dagger)
+  \left(Y_\nu Y_\nu^\dagger \right)^T M_\nu \right] 
+ M_\nu \left(q_T T + q_{g_1} g_1^2 + q_{g_2} g_2^2 + 
q_{g_3} g_3^2 + q_{\lambda} \lambda \right) \; , \qquad
\eeqa
where $T$ is given by Eq.~(\ref{T}). All of $q_i$s are expected to be
of ${\cal O}(1)$. As already discussed, the trace term $T$ can appear
only through Higgs interactions, and so it cannot be present here
since $M_\nu$ does not couple to Higgs rendering $q_T=0$. Moreover,
since the added lepton fields $\nu_R^i$ are singlets under U(1)$_Y$ and
$\su(3)_C$, $q_{g_1} = q_{g_3} =0$. At this order, $\lambda$ dependence
cannot appear either making $q_{\lambda}=0$. So we are left with
\beqa
\dot M_\nu &=& \frac{q_1}{2} \left[ M_\nu \left( Y_\nu Y_\nu^\dagger \right)  
+  \left(Y_\nu Y_\nu^\dagger \right)^T M_\nu \right] 
+ q_{g_2} g_2^2 M_\nu \; .
\label{RG-Mnu}
\eeqa
Here, we keep the $g_2^2$ dependence to get the general form of $\dot M_\nu$ 
for right-handed neutrino extended models with $\Glf$ flavor symmetry.
For MLFV-ex, where the right-handed neutrinos 
are singlets under $\su(2)_L$, we have $q_{g_2}=0$.
It should also be noted that if we use the universality of $M_\nu$ as 
an initial condition in Eq.~(\ref{RG-Mnu}), when $\Glf$ is broken by the small 
background values of the spurion field $Y_\nu$, the universality 
of the Majorana mass matrix is also broken as $Y_\nu$ has 
non-zero off-diagonal entries in general. However, the breaking 
is small and we can still consider $\Glf$ as the flavor symmetry 
of the theory in the massless lepton limit and perform the 
spurion analysis.

Let us now consider the term containing $M_\nu^\dagger$, that 
involves the left-handed fields. Writing the indices
explicitly, for a general $M_\nu$ matrix, we get that the mass term
associated with the left-handed fields is ${(M_\nu^\ast)}_{\alpha
\beta}$, instead of ${(M_\nu)}^{\alpha \beta}$ for the right-handed 
fields, and thus the allowed terms are
\beqa
T_\nu^\prime M_\nu &=& (1,1,8) \otimes (1,1,6) \ni (1,1,6) \; , \\
\tr[Y_e^\dagger Y_e] M_\nu &=& (1,1,1) \otimes (1,1,6) = (1,1,6) \; , \\
\tr[Y_\nu^\dagger Y_\nu] M_\nu &=& (1,1,1) \otimes (1,1,6) = (1,1,6) \; .
\eeqa
Hence after symmetrization the evolution equation 
of the right-handed neutrino mass has a Dirac structure and is 
given by
\beqa
\dot M_\nu &=& \frac{q_1}{2} \left[ \left( M_\nu \left( Y_\nu Y_\nu^\dagger \right)  
+  \left(Y_\nu Y_\nu^\dagger \right)^T M_\nu \right) P_R
+ \left( \left( Y_\nu Y_\nu^\dagger \right) M_\nu  
+ M_\nu \left(Y_\nu Y_\nu^\dagger \right)^T \right) P_L \right]
+ q_{g_2} g_2^2 M_\nu \; . \nn \\
\label{Mnu-complete}
\eeqa
Eq.~(\ref{Mnu-complete}) is the most general form of $M_\nu$ 
evolution, as obtained by loop diagram calculations in 
\cite{Type-I-rg,triplet-RG}.

%%%%%%%%%%%%%%%%%%%%%%%%%%%%%%%%%%%%%%%%%%%%%%%%%%%%%%%%%%%%%%%%%%
%%%%%%%%%%%%%%%%%%%%%%%%%%%%%%%%%%%%%%%%%%%%%%%%%%%%%%%%%%%%%%%%%%
\subsection{RG evolution of the left-handed Majorana mass $m_\nu$}
%%%%%%%%%%%%%%%%%%%%%%%%%%%%%%%%%%%%%%%%%%%%%%%%%%%%%%%%%%%%%%%%%%
%%%%%%%%%%%%%%%%%%%%%%%%%%%%%%%%%%%%%%%%%%%%%%%%%%%%%%%%%%%%%%%%%%

At energy scales above $M_R$, the light left-handed neutrino mass
$m_\nu$ is generated through the seesaw relation and hence the RG
evolution of $m_\nu$ will be obtained through that of $Y_\nu$ and
$M_\nu$, as given in Eqs.~(\ref{Ynu}) and (\ref{RG-Mnu}). Using the
seesaw relation given in Eq.~(\ref{mnu-Glf}) and considering the fact
that ${(M_\nu^{-1})}^{\alpha \gamma}{(M_\nu)}_{\gamma \beta} =
{\delta^\alpha}_\beta$, we see that to get the RG evolution equation
for ${(m_\nu)}^{ij}$, $i,j$ being the $\su(2)_{l_L}$ indices, 
one needs the evolution of ${(M_\nu})_{\alpha \beta}$, 
i.e. the left-chiral projection of the RG evolution of $M_\nu$, which
can be read off from Eq.~(\ref{Mnu-complete}). Finally, the evolution
equation for $m_\nu$ is given by
\beqa
\dot m_\nu &=& m_\nu P + P^T m_\nu + p m_\nu \; , 
\label{RG-mnu}
\eeqa
where
\beqa
P &=& b_1 Y_e^\dagger Y_e + \left( b_2 - \frac{q_1}{2} \right) Y_\nu^\dagger Y_\nu \; , 
\label{RG-mnu-P} \\
p &=& 2 \left( b_T T + b_{g_1} g_1^2 + b_{g_2} g_2^2  \right) - q_{g_2} g_2^2  \; .
\label{RG-mnu-p}
\eeqa
Note that the RHS of the equation is symmetric under $\su(3)_{l_L}$, 
as required. All of $b_{1,2}$, $b_T$, $b_{g_{1,2}}$ and
$q_1$ are given below for the cases of Type-I and Type-III seesaw in
SM and MSSM.

%%%%%%%%%%%%%%%%%%%%%%%%%%%%%%%%%%%%%%%%%%%%%%%%%%%%%%%%%
%%%%%%%%%%%%%%%%%%%%%%%%%%%%%%%%%%%%%%%%%%%%%%%%%%%%%%%%%
\subsection{RG evolution at energies below $M_R$}
%%%%%%%%%%%%%%%%%%%%%%%%%%%%%%%%%%%%%%%%%%%%%%%%%%%%%%%%%
%%%%%%%%%%%%%%%%%%%%%%%%%%%%%%%%%%%%%%%%%%%%%%%%%%%%%%%%%

To complete the discussion of RG evolution of the different quantities
that are needed in order to have a complete description of all
leptonic parameters at all energy scales, we now construct the RG
evolution equations for $\mu < M_R$. In this regime the flavor
symmetry is
\beq
\Glf^\prime \equiv \su(3)_{l_L} \otimes \su(3)_{e_R} \; ,
\eeq 
and the Yukawa coupling, $Y_e(\bar 3, 3)$, and the left-handed Majorana
mass, $m_\nu(6,1)$, are the only spurion fields. The RG evolution
equation for $Y_e$ can be obtained following the procedure given in
the last subsection to be
\beqa
\dot Y_e &=& Y_e \left( a_1 Y_e^\dagger Y_e  + a_{T} T^\prime + a_{g_1} g_1^2 + a_{g_2} g_2^2 \right) \; ,
\eeqa
where 
\beq
T^\prime \equiv \tr[Y_e^\dagger Y_e] + 3\tr[Y_U^\dagger Y_U] + 3\tr[Y_D^\dagger Y_D] \;,
\label{Tprime}
\eeq
and $a_i$s are expected to be of ${\cal O}(1)$ as before.

In the low energy regime $m_\nu$ is an effective neutrino mass
operator and its RG evolution is not given by Eq.~(\ref{RG-mnu}). To 
determine the structure of the RG evolution equation for the
left-handed Majorana mass $m_\nu$, we proceed in the same way as 
in case of $M_\nu$, keeping in mind the change in the chirality.
Table~\ref{tab:trans} and the transformation rule in
Eq.~(\ref{su3-mnu}) can be used to determine the allowed combinations
of $m_\nu$ and $Y_e$ that can appear on the RHS of $\dot m_\nu$ and
those are $m_\nu T_e$ and $m_\nu \tr[Y_e^\dagger Y_e]$. Symmetrized
over the $\su(3)_{l_L}$ indices, the most general form of $\dot
m_\nu$, keeping 1-loop spurion contributions, is
\beqa
\dot m_\nu &=& \frac{p}{2} \left( m_\nu T_e + (m_\nu T_e)^T \right)
+ p_e\tr[Y_e^\dagger Y_e]  \nn \\
&+& m_\nu \left( p_U \tr[Y_U^\dagger Y_U] + p_D \tr[Y_D^\dagger Y_D] 
+ p_{g_1} g_1^2 + p_{g_2} g_2^2 + p_\lambda \lambda \right) \; ,
\eeqa
which can be simplified to
\beqa
\dot m_\nu &=&  \frac{p_1}{2} \left( m_\nu Y_e^\dagger Y_e + (Y_e^\dagger Y_e)^T m_\nu \right)+ 
m_\nu ( p_T T^\prime + p_{g_1} g_1^2 + p_{g_2} g_2^2 + 
p_\lambda \lambda )\; , \quad
\eeqa
where $T^\prime$ has been defined in Eq.~(\ref{Tprime}). 
As before, we have considered the $\su(3)_C$ charges of the 
quarks in fixing $p_{U,D}$ and writing $T^\prime$.
Here $p_i$s are the ${\cal O}(1)$ numbers and we have 
used the fact that $m_\nu$ is symmetric under $\su(3)_{l_L}$. 

%%%%%%%%%%%%%%%%%%%%%%%%%%%%%%%%%%%%%%%%%%%%%%%%%%%%%%%%%%%%%%%%%%%%%
%%%%%%%%%%%%%%%%%%%%%%%%%%%%%%%%%%%%%%%%%%%%%%%%%%%%%%%%%%%%%%%%%%%%%
\section{Results}
\label{sec:results}
%%%%%%%%%%%%%%%%%%%%%%%%%%%%%%%%%%%%%%%%%%%%%%%%%%%%%%%%%%%%%%%%%%%%%
%%%%%%%%%%%%%%%%%%%%%%%%%%%%%%%%%%%%%%%%%%%%%%%%%%%%%%%%%%%%%%%%%%%%%

To illustrate the RGEs obtained in Section~\ref{sec:RG} using spurion analysis, 
we compare the coefficients with the evolution
equations obtained by exact calculations in four different
models. These models are the extended SM and MSSM, where the
right-handed neutrinos can be singlets (Type-I
seesaw~\cite{Type-I-rg,MSSM-1st-order, MSSM-2nd-order}) or triplets
(Type-III seesaw~\cite{triplet-RG}).

%%%%%%%%%%%%%%%%%%%%%%%%%%%%%%%%%%%%%%%%%%%%%%%%%%%%%%%%%%%%%%%%%%%%
%%%%%%%%%%%%%%%%%%%%%%%%%%%%%%%%%%%%%%%%%%%%%%%%%%%%%%%%%%%%%%%%%%%%
\subsection{Right-handed neutrino extended SM}
%%%%%%%%%%%%%%%%%%%%%%%%%%%%%%%%%%%%%%%%%%%%%%%%%%%%%%%%%%%%%%%%%%%%
%%%%%%%%%%%%%%%%%%%%%%%%%%%%%%%%%%%%%%%%%%%%%%%%%%%%%%%%%%%%%%%%%%%%

Let us first consider the case of the SM extended with three
right-handed neutrinos. There can be only two possibilities: the 
first option is when the right-handed neutrinos are singlets under 
the gauge group which is
known as Type-I seesaw. The other option, known as Type-III seesaw,
is when the neutrinos are triplets under $\su(2)_L$ and singlet under
the remaining $\su(3)_C \times U(1)_Y$. Note that for
Type-II seesaw \cite{type-II} as well as Inverse seesaw
\cite{inverse-seesaw}, the flavor group and the spurions present 
in the theory are not identical to the above cases 
and cannot be treated as a realization of the case discussed here.

In the general case of Type-I and Type-III seesaw, 
each of the right-handed neutrinos can be expressed as
\beq
\nu_R \equiv \sum_{a=1}^N \nu_R^a G^a \; , 
\eeq
with 
\beqa
\barr{llll}
G^a \equiv {\mathbbm I} \; , &  N = 1  &  & {\text{for Type-I seesaw}} \; , \\
G^a \equiv \sigma^a  \; , &  N = 3 &  & {\text{for Type-III seesaw}} \; ,
\earr
\label{Ga}
\eeqa
where $\sigma^a$ represent the Pauli matrices. Note that we work in
three different spaces. The flavor index, $f=e,\mu,\tau$, is
suppressed. There is also the internal SU(2) index of the Pauli
matrices that we suppress here and in the rest of the paper. In the
following we often get quantities that are universal in that
index. Last, the explicit index $a$ that runs from 1 to $N$.

With the above definition, we can write 
$r \equiv a_4/a_3$ in Eq.~(\ref{T}) as 
\beq
r = \sum_{a=1}^N \epsilon^T G^a G^a \epsilon = (1, 3) \; , 
\label{r-SM}
\eeq
where $\epsilon \equiv i \sigma^2$. The two numbers in the
parenthesis are the values in Type-I and Type-III seesaws, 
and are universal in the SU(2) spaces.

The quantities that appear in the coefficients of $\dot Y_e$, 
$\dot Y_\nu$ and $\dot M_\nu$, and depend on the 
representation of the right-handed neutrinos are
\beqa
\alpha_1&=& \sum_{a=1}^N  G^a G^a=(1,3)\; , \\
\alpha_2&=& \tr[G^a G^a]=(2,2)\; , \label{alpha2} \\
\alpha_3 &=& \sum_{a=1}^N G^{aT} \epsilon G^a \epsilon =(-1,3) \; , \label{alpha3}\\
\alpha_4 &=& {\left( \epsilon^T G^a \right)}^T 
{\left( \epsilon^T G^a \right)}^{-1} = ( -1, 1)\; , \\
\alpha_5 &=& \sum_{a,b=1}^N 
\left( i \varepsilon^{bac} G^{aT} \epsilon^T G^b \right)
{\left( \epsilon^T G^c \right)}^{-1}  = ( 0, -2)\; ,
\eeqa
where $\varepsilon^{bac}$ is the completely anti-symmetric tensor in
$\su(2)$ indices and no summation convention has been used.

Let us now discuss the origin of $\alpha_i$s. $\alpha_1$ comes from
the self-energy correction of $l_L$, while $\alpha_2$ 
appears in the self-energy correction of $\nu_R$. $\alpha_3$ comes in
the correction of the vertex containing $Y_e$, while $\alpha_4$ is
present in the correction of the $Y_\nu$ vertex. $\alpha_5$ appears
in the vertex correction of $Y_\nu$ because of $\su(2)_L$
interactions. In the case of right-handed neutrino extended SM,
self-energy, mass and vertex corrections contribute to the running of
the Yukawa couplings $Y_{e,\nu}$. Hence, $\alpha_1$ is expected to
contribute to both $\dot Y_e$ and $\dot Y_\nu$, while $\dot Y_e$
should contain $\alpha_3$ as well. $\alpha_4$ and $\alpha_5$ must
appear in $\dot Y_\nu$. As already discussed, these quantities do not
appear in $\dot m_\nu$ in the regime $\mu < M_R$, since the
right-handed neutrinos are already decoupled.

Let us now consider the coefficients $a_{1,2}$, $a_T$ and $a_{g_1,g_2}$ 
arising in $\dot Y_e$ in Eq.~(\ref{Ye}). 
Collecting all the contributions, we get the 
coefficients in Eq.~(\ref{Ye}) to be~\cite{Type-I-rg,triplet-RG}
\beq
a_1 = \frac{3}{2} \; , \quad
a_2 = \frac{\alpha_1}{2} + 2 \alpha_3 \; , \label{a2-sm} \quad
a_T = 1 \; , \quad
a_{g_1} = \left( -\frac{3}{4} -3 \right) \times \frac{3}{5} \; , \quad
a_{g_2} = -3 \times \frac{3}{4} \; . 
\eeq
The first term in $a_{g_1}$ arises through the self-energy 
correction of Higgs field $\phi$, which also contributes 
to $a_{g_2}$. Here we have used GUT normalization 
for U(1)$_Y$ charges and hence a factor of $(3/5)$ comes with $g_1^2$.
The coefficients appearing in the RG evolution equation of $Y_\nu$ in
Eq.~(\ref{Ynu}) can also be obtained in a similar way and we
have~\cite{Type-I-rg,triplet-RG}
\beq
b_1 = \frac{1}{2} + 2\alpha_4 \; , \label{b1-sm}\quad
b_2 = \frac{1}{2} \left( \alpha_1 + \alpha_2 \right) \; , \quad
b_T = 1 \; , \quad
b_{g_1} = - \frac{3}{4} \times \frac{3}{5}\; , \quad
b_{g_2} = -3 \times \frac{3}{4} + 3 \alpha_5  \; .
\eeq
\begin{table}
\begin{center}
\begin{tabular}{|c|c|c|c|c|c|}
\hline
& & \multicolumn{2}{|c|}{SM} & \multicolumn{2}{|c|}{MSSM}\\
\hline 
& & Type-I & Type-III & Type-I & Type-III \\
\hline
$T (T_U)$ & $r (r^\prime)$ & 1 & 3  & 1 & 3\\
\hline
& $a_1$ & $3/2$ & $3/2$ & $3$ & $3$ \\
& $a_2$ & $-3/2$ & $15/2$  & $1$ & $3$ \\
$\dot Y_e$ & $a_T$ & $1$ & $1$  & $1$ & $1$\\
& $a_{g_1}$ & $-9/4$ & $-9/4$ & $-9/5$ & $-9/5$ \\
& $a_{g_2}$ & $-9/4$ & $-9/4$ & $-3$ & $-3$ \\
\hline 
& $b_1$ & $-3/2$ & 5/2  & 1 & 1 \\
& $b_2$ & 3/2 & 5/2  & 3 & 5\\
$\dot Y_\nu$ & $b_T$ & 1 & 1  & 1 & 1 \\
& $b_{g_1}$ & $-9/20$ & $-9/20$  & $-3/5$ & $-3/5$ \\
& $b_{g_2}$ & $-9/4$ & $-33/4$ & $-3$ & $-7$ \\
\hline
$\dot M_\nu$ & $q_1$ & 2 & 2  & 4 & 4\\
& $q_{g_2}$ & 0 & $-12$  & 0 & $-8$ \\
\hline
& $p_1$ & \multicolumn{2}{|c|}{$-3$} & \multicolumn{2}{|c|}{$-3$}\\
$\dot m_\nu$ & $p_T$ & \multicolumn{2}{|c|}{2}  & \multicolumn{2}{|c|}{2}\\
$(\mu < M_R)$ & $p_{g_1}$ & \multicolumn{2}{|c|}{0} & \multicolumn{2}{|c|}{$-6/5$} \\
& $p_{g_2}$ & \multicolumn{2}{|c|}{$-3$} & \multicolumn{2}{|c|}{$-6$} \\
& $p_\lambda$ & \multicolumn{2}{|c|}{1} & \multicolumn{2}{|c|}{0} \\
\hline 
\end{tabular}
\end{center}
\caption{Coefficients appearing in the RG evolution of 
$Y_e$, $Y_\nu$, $M_\nu$ and $m_\nu$ in the SM and MSSM, 
in case of Type-I and Type-III seesaw 
\cite{Type-I-rg,triplet-RG,MSSM-1st-order, MSSM-2nd-order}. 
For the extended MSSM, $a_T$ is the coefficient of $T_D$, 
while $b_T$ and $p_T$ are of $T_U$ and $T_U^\prime$, respectively.
\label{tab:1loop-coeff}}
\end{table}
The values of $a_i$ and $b_i$ in Type-I and Type-III 
seesaw scenarios are tabulated in Table~\ref{tab:1loop-coeff}.
As can be seen from the table, for Type-I seesaw in the extended SM 
model the coefficients are ${\cal O}(1)$ numbers, as expected. 
In the case of the Type-III seesaw, we see that there are numbers
which are larger than ${\cal O}(1)$, 
for example $a_2$ and $b_{g_2}$. Let us now try to understand the 
origin of these large numbers. 
The largest contribution to $a_2$ comes from the $\alpha_3$ in
Eq.~(\ref{a2-sm}), which arises through the vertex correction due to
right-handed triplets and a factor of three is expected. Thus, the
relevant number which we expect to be of ${\cal O}(1)$ is $(a_2/3)$.
Moreover, the right-handed neutrino triplets have interactions with
the $\su(2)_L$ gauge bosons over the singlets, and so we expect
$b_{g_2}$ in the Type-III case to have a factor of six over $b_{g_2}$ in
Type-I.

Let us now discuss the coefficients $q_1$ and $q_{g_2}$
appearing in the running of $M_\nu$. The coefficients are 
given by
\beq
q_1 = \alpha_2 \; ,\qquad  q_{g_2} =  (0,-12)\;,
\eeq
where $\alpha_2$ is defined in Eq.~(\ref{alpha2}) and is of ${\cal
O}(1)$. For Type-I seesaw, the right-handed neutrinos are singlets of
$\su(2)_L$ and so $q_{g_2} = 0$, while for Type-III seesaw one gets
by exact calculations \cite{triplet-RG} $q_{g_2} = -12$ and
($q_{g_2}/6$) is of ${\cal O}(1)$, as discussed earlier.

Last, we consider the evolution of the effective left-handed Majorana 
neutrino mass $m_\nu$ in the energy scales $\mu < M_R$. 
In this energy regime, the evolution equations are the same for all
the different seesaws, since we are considering an effective
theory. However they will depend on the underlying theory, which is 
the SM in this case. The values of different $p_i$s are given in
Table~\ref{tab:1loop-coeff} and are of ${\cal O}(1)$ as anticipated.

Note that explicit 1-loop calculations show that $p_{g_1} = 0$. We
were unable to find an explanation based on symmetry considerations
and hence we think it is accidental. We expect $g_1^2$ dependent terms 
to emerge at 2-loop.

%%%%%%%%%%%%%%%%%%%%%%%%%%%%%%%%%%%%%%%%%%%%%%%%%%%%%%%%%%%%%%%%%%%%%%%%%%
%%%%%%%%%%%%%%%%%%%%%%%%%%%%%%%%%%%%%%%%%%%%%%%%%%%%%%%%%%%%%%%%%%%%%%%%%%
\subsection{Right-handed neutrino extended MSSM}
%%%%%%%%%%%%%%%%%%%%%%%%%%%%%%%%%%%%%%%%%%%%%%%%%%%%%%%%%%%%%%%%%%%%%%%%%%
%%%%%%%%%%%%%%%%%%%%%%%%%%%%%%%%%%%%%%%%%%%%%%%%%%%%%%%%%%%%%%%%%%%%%%%%%%

We now consider the case of the MSSM extended by three right-handed
neutrinos. Our formalism is applicable in this case 
as well, since the flavor structure of the MSSM is identical to that
of the SM. But the Higgs sector of MSSM is different.
One of the Higgses, $H_U$, couples to leptons through the
Yukawa coupling $Y_U$ to give rise to the up-type lepton masses, while
the other Higgs, $H_D$, is responsible for the down-type lepton masses
through the Yukawa coupling $Y_D$. Hence there are two types of trace
terms. The first is $T_U$ which is a combination of $\tr [Y_\nu^\dagger
Y_\nu]$ and $\tr[Y_U^\dagger Y_U]$. The other one is $T_D$, a
combination of $\tr [Y_e^\dagger Y_e]$ and $\tr[ Y_D^\dagger Y_D]$.
We define the trace terms as
\beqa
T_U &=& r^\prime \tr [Y_\nu^\dagger Y_\nu] + 3 \tr[Y_U^\dagger Y_U] \;
, \label{TUMSSM} \\
T_D &=& \tr [Y_e^\dagger Y_e ]+ 3 \tr[Y_D^\dagger Y_D] \; , \label{TDMSSM}
\eeqa
where
\beq
r^\prime \equiv \sum_{i=1}^N \left( \epsilon G^i \right)^\ast \left(
\epsilon G^i \right)^T = ( 1, 3 ) 
\label{r-prime}
\eeq
is a quantity, similar to $r$ defined in Eq.~(\ref{r-SM}) in the SM, that
depends on the transformation of the right-handed neutrinos under the
gauge group. The two numbers in the parenthesis are the values in
Type-I and Type-III seesaw scenarios. 
As before, $r^\prime$ is universal in SU(2) spaces and 
we write down the universality constant only.

Let us now define the quantities that contribute to the evolution of 
$Y_e$, $Y_\nu$ and $M_\nu$ in Type-I and Type-III seesaws and depend on the 
gauge group representations of the right-handed neutrinos:
\beqa
\alpha^\prime_1 &=& \sum_{a=1}^N \left( \epsilon G^a \right)^\dagger \left( \epsilon G^a \right) = 
(1, 3) \; ,
\label{alpha1-prime}\\
\alpha^\prime_2 &=&  \tr[\left( \epsilon G^a \right)^\dagger \left( \epsilon G^a \right) ]= (2, 2) \; ,
\label{alpha2-prime} \\
C_2 &=& (0,2) \; .
\label{casimir}
\eeqa
$C_2$ is the quadratic Casimir for the irreducible representation
${\cal R}$ of $\su(2)_L$ in which the right-handed neutrinos $\nu_R^i$
reside. For Type-I seesaw $C_2= 0$, while for Type-III seesaw the
right-handed fields are in the adjoint representation of $\su(2)_L$
and hence $C_2 = 2$. RG evolution of Yukawas and 
masses in Type-III seesaw with MSSM as the underlying theory 
has not been computed before We give some details of the 
calculation in Appendix~\ref{app:MSSM-typeIII}.

Let us now write down the coefficients involved in $\dot Y_e$ in Eq.~(\ref{Ye}).
\beq
a_1 = 3 \; , \quad
a_2 = \alpha^\prime_1 \; , \quad
a_{T} = 1 \; , \quad
a_{g_1} = -3 \times \frac{ 3}{5} \; , \quad
a_{g_2} = -3  \; .
\eeq
We see that in the MSSM, as in the case of the SM, only $a_2$, the
coefficient of $Y_\nu^\dagger Y_\nu$, depends on whether the seesaw
is Type-I or Type-III. For the case of $\dot Y_\nu$, the
coefficients appearing in Eq.~(\ref{Ynu}) are
\beq
b_1 = 1 \; , \label{b1-mssm} \quad
b_2 = \alpha^\prime_1 + \alpha^\prime_2 \; , \quad
b_{T}= 1 \; , \quad
b_{g_1} = - \frac{3}{5} \; , \quad
b_{g_2} = -3 - 2 C_2  \; .
\eeq
Comparing the expressions of $b_1$ in the SM and the MSSM, 
in Eqs.~(\ref{b1-sm}) and (\ref{b1-mssm}), we see that 
in the SM $b_1$ receives a contribution that 
depends on the right-handed neutrinos, which is absent in MSSM. This 
is to be attributed to the non-renormalization theorem due to which 
only the wavefunction renormalizations are responsible for the RG 
evolution of the quantities in MSSM and the mass and vertex corrections 
do not contribute. The absence of any vertex renormalization contribution 
makes $b_1$ independent of the right-handed neutrino fields in MSSM.
The values of $a_i$ and $b_i$ in the two seesaw types are given in
Table~\ref{tab:1loop-coeff}.

From Table~\ref{tab:1loop-coeff} it is seen that for Type-I seesaw 
scenario, all the numbers are of ${\cal O}(1)$ and consistent with 
prediction from spurion analysis.
However, for Type-III seesaw both $b_2$ and $b_{g_2}$ are large numbers, 
the large contribution emerging from the wavefunction renormalization of 
the superfields $l$ and $\nu$ respectively.

Next, we move to the case of the right-handed Majorana mass $M_\nu$. 
The coefficients are 
\beq
q_1 = 2 \alpha_2^\prime \; , \quad
q_{g_2} = -4 C_2 \; ,
\eeq
where $\alpha^\prime_2$ and $C_2$ have already been defined in
Eqs.~(\ref{alpha2-prime}) and (\ref{casimir}) respectively.
Values of $q_1$ and $q_{g_2}$ in the two types of seesaw scenarios 
are listed in Table~\ref{tab:1loop-coeff}. As expected, $q_{g_2}=0$ 
and $q_1$ is of ${\cal O}(1)$ in Type-I seesaw, while 
for Type-III seesaw $q_1$ and $(q_{g_2}/6)$ are ${\cal O}(1)$ numbers.

For energies $\mu < M_R$, evolution of the left-handed neutrino 
mass $m_\nu$ is the same in both Type-I and Type-III seesaws and 
the values of the coefficients \cite{MSSM-1st-order, MSSM-2nd-order} 
are quoted in Table~\ref{tab:1loop-coeff}. 
Note that the accidental cancellation seen in the SM case, 
$p_{g_1}=0$, does not happen in the MSSM.
The trace term appearing in this case is 
$T^\prime \to T_U^\prime = \tr[3 Y_U^\dagger Y_U]$, since in the 
high energy theory only $H_U$ interacts with $\nu$.
The Higgs self-coupling term with coefficient $p_\lambda$ does not exist 
in this scenario.

The above comparison shows that the method of spurion analysis gives
the form of the RG evolution equations. Of course, working in a generic
effective field theory we never expect to get the exact values of the
${\cal O}(1)$ numbers, which depend on the specific details of the
model. One can use this same technique to get the evolution equations
at second order. Calculation of evolution equations at 2-loop and
comparison with the existing results obtained by loop calculations is
given in Appendix~\ref{app:2loop}.

%%%%%%%%%%%%%%%%%%%%%%%%%%%%%%%%%%%%%%%%%%%%%%%%%%%%%%%%%
%%%%%%%%%%%%%%%%%%%%%%%%%%%%%%%%%%%%%%%%%%%%%%%%%%%%%%%%%
\section{Breaking degeneracy of $M_\nu$ and leptogenesis}
\label{sec:example}
%%%%%%%%%%%%%%%%%%%%%%%%%%%%%%%%%%%%%%%%%%%%%%%%%%%%%%%%%
%%%%%%%%%%%%%%%%%%%%%%%%%%%%%%%%%%%%%%%%%%%%%%%%%%%%%%%%%

In this section, we study effects related to the breaking of the
universality of $M_\nu$. This breaking is important in the context of
leptogenesis. It has been studied in detail
in~\cite{MLFV-leptogenesis-Isidori} where the mass degeneracy is
removed by appropriate combinations of spurions transforming as $(1,
1, 6)$ under $\Glf$. Here we compare their results of 
explicit breaking with the effects generated through RG evolution.

We start with the case of degeneracy breaking by RG evolution. 
For this purpose, 
writing down the evolution equation for a component of $M_\nu$ from
Eq.~(\ref{RG-Mnu}) we get
\beqa
\left( \dot M_\nu \right)_{ij} &=& 
\frac{q_1}{2} \left[ \left( M_\nu \right)_{ik} \left( Y_\nu Y_\nu^\dagger \right)_{kj} + 
\left( Y_\nu Y_\nu^\dagger \right)_{ki} \left( M_\nu \right)_{kj} \right] + 
q_{g_2} g_2^2 \left( M_\nu \right)_{ij} \; . 
\eeqa
Using universal-mass initial condition, $\left( M_\nu \right)_{ij} = M_R
\delta_{ij}$, one gets the final eigen-values of $M_\nu$ after RG
running to be non-degenerate. The specific value of breaking depends
on the values of $q_1$, $q_{g_2}$ as well as the RG evolution of the
spurion field $Y_\nu$ and its background value, and thus on the
underlying theory considered.

Next, we study degeneracy breaking at the high scale using spurion 
techniques. To the lowest order in the spurion fields $Y_{e,\nu}$, the 
final Majorana mass matrix $M_\nu^F$ is written as
\beq
M_\nu^F = M_\nu + \sum_n c_n \delta M_\nu^{(n)} \; , 
\label{break-Isidori}
\eeq
where $M_\nu = M_R {\mathbbm I}$ is the universal mass matrix given in Eq.~(\ref{Mnu-uni}) and
\beqa
\delta M_\nu^{(11)} &=& M_R \left( Y_\nu Y_\nu^\dagger + (Y_\nu Y_\nu^\dagger)^T \right) \; , \nn \\
\delta M_\nu^{(21)} &=& M_R \left( Y_\nu Y_\nu^\dagger Y_\nu Y_\nu^\dagger  
+ (Y_\nu Y_\nu^\dagger Y_\nu Y_\nu^\dagger)^T \right) \; , \nn \\
\delta M_\nu^{(22)} &=& M_R \left( Y_\nu Y_\nu^\dagger (Y_\nu Y_\nu^\dagger)^T \right) \; , \nn \\
\delta M_\nu^{(23)} &=& M_R \left( (Y_\nu Y_\nu^\dagger)^T Y_\nu Y_\nu^\dagger  \right) \; , \nn \\
\delta M_\nu^{(24)} &=& M_R \left( Y_\nu Y_e^\dagger Y_e Y_\nu^\dagger + (Y_\nu Y_e^\dagger Y_e Y_\nu^\dagger)^T \right) \; ,
\eeqa
considering terms containing up to four spurions. As discussed in
\cite{MLFV-leptogenesis-Isidori}, values of $c_{n}$ depends on
dynamical properties: if the Yukawa corrections are generated within a
perturbative regime, as is the case for RG evolution, $c_n$ decreases
according to the power of Yukawa matrices, for example, in a standard
loop-expansion one should have $c_{11} \sim g_{\rm eff}^2/(4 \pi)^2$ and
then $c_{2i} \sim c_{11}^2$ and so on. One cannot exclude a priori a
strong-interaction regime where $c_n \sim {\cal O}(1)$, for all
$n$. But even in the case of strong-interaction, the series in
Eq.~(\ref{break-Isidori}) is expected to be dominated by the first few
terms as the background values of the spurions $Y_{e, \nu}$ are small. 
In this paper, we consider the perturbative regime of explicit breaking 
only.

In Ref.~\cite{MLFV-leptogenesis-Isidori} it is shown that the amount
of mass degeneracy breaking is important in the context of
leptogenesis. In the rest of this section we consider the two sources
of breaking and study the pattern of mass universality breaking and
its effect on leptogenesis. We briefly describe the parametrization of
the Yukawa $Y_\nu$ following~\cite{MLFV-leptogenesis-Isidori}. We
choose to work in the basis where $Y_e$ is diagonal. Then the neutrino
mass matrix is given as
\beqa
m_\nu = U_{\rm PMNS}^\ast m_\nu^{\rm diag} U_{\rm PMNS}^\dagger \; , 
\eeqa
where 
\beq 
m_\nu^{\rm diag} = {\rm diag}(m_1, m_2, m_3)
\eeq
and $U_{\rm PMNS}$ is the unitary matrix that diagonalizes $m_\nu$.
In this basis, the most general form of $Y_\nu$ is given by the
Casas-Ibarra parametrization~\cite{casas-ibarra}:
\beqa
Y_\nu &=& \frac{1}{v} M_\nu^{1/2} R {\left(m_\nu^{\rm diag}\right)}^{1/2} U_{\rm PMNS}^\dagger 
= \frac{\sqrt{M_R}}{v} R {\left(m_\nu^{\rm diag}\right)}^{1/2} U_{\rm PMNS}^\dagger\; ,
\eeqa
where $R$ is a complex orthogonal matrix parametrized by six real
quantities. We write $R = O H$, where $O$ is a real orthogonal matrix
and $H$ is complex orthogonal hermitian matrix and thus each $O$ and
$H$ contains three real parameters. Since $O \in O(3)_{\nu_R}$, and
$O(3)_{\nu_R}$ is a symmetry of the theory independent of any
assumption on CP properties, we can choose $O \equiv {\mathbbm I}$ to
get $R = H$. Thus finally
\beqa
Y_\nu &=& \frac{\sqrt{M_R}}{v}  H {\left(m_\nu^{\rm diag}\right)}^{1/2} U_{\rm PMNS}^\dagger \; .
\eeqa
In the CP conserving limit, $H = {\mathbbm I}$. The CP violating nature of $H$ is clear in the 
following parametrization~\cite{H-param}:
\beqa
H = e^{i \Phi} = {\mathbbm I} - \frac{\cosh{\rho} -1}{\rho^2} \Phi^2 + i \frac{\sinh{\rho}}{\rho} \Phi \; ,
\label{H}
\eeqa
where
\beqa
\rho = \sqrt{\varphi_1^2 + \varphi_2^2 + \varphi_3^2} \; , \qquad 
{\rm and} \qquad 
\Phi &=& \left(
\barr{ccc}
0 & \varphi_1 & \varphi_2 \\
-\varphi_1 & 0 & \varphi_3 \\
-\varphi_2 & -\varphi_3 & 0
\earr
\right) \; .
\label{phi-H-param}
\eeqa

\begin{figure}[t!]
\begin{center}
\includegraphics[scale=0.7]{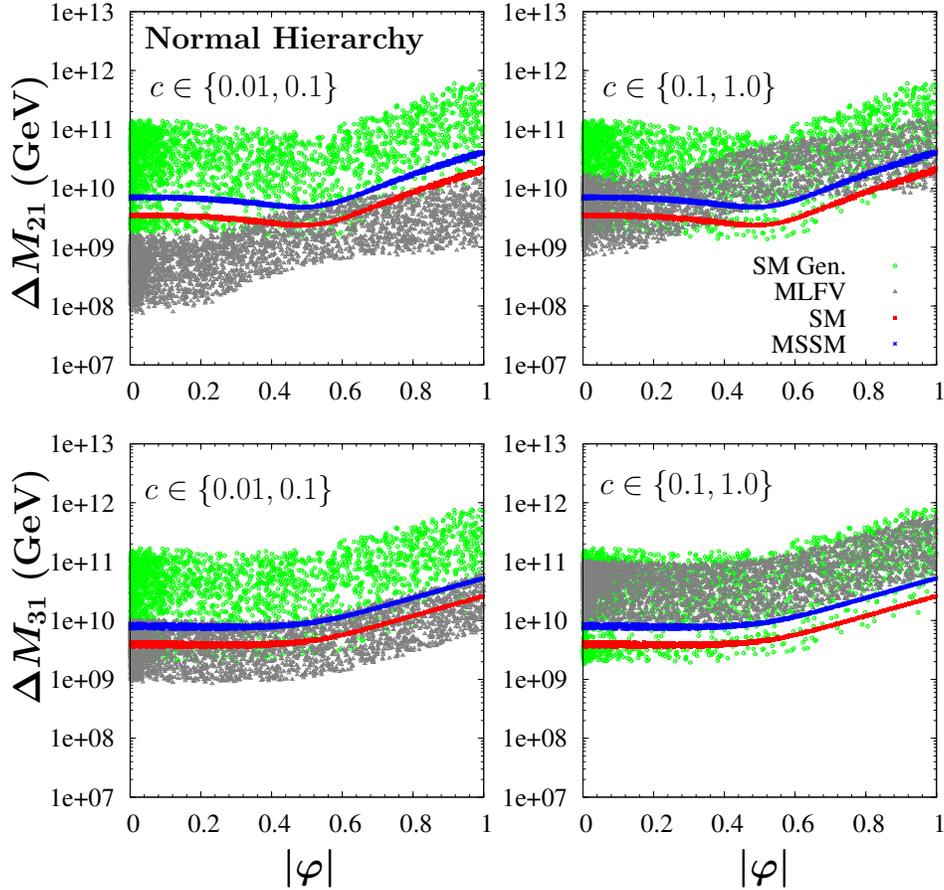}
\caption{Majorana mass splittings as a function 
of $|\varphi|$ for normal neutrino mass hierarchy. 
We have defined $\Delta M_{i1} = M_i - M_1$. 
The green (light gray in black and white) dots show 
`SM Gen.', and dark gray dots are for `MLFV'. The red (lower) 
and blue (upper) dotted lines correspond to SM and MSSM 
respectively.
\label{fig:Type-I-normal}}
\end{center}
\end{figure}
Let us now proceed to the numeric example. In the generic case of
\cite{MLFV-leptogenesis-Isidori}, the breaking depends on the choice
of $c_n$s, while in case of RG evolution we need to specify the
underlying theory (for example SM or MSSM and also Type-I or
Type-III). In both cases, the mass-splitting of the right-handed
neutrinos depends on $\Phi$ as well as the neutrino masses and mixing 
parameters through $Y_\nu$. 
For the purpose of illustration we choose $M_R = 10^{13}$ GeV, 
and $\varphi_1 = \varphi_2 = \varphi_3 =\varphi$, and then 
consider the range $10^{-3} \leq |\varphi| \leq 1$. 
The neutrino mass-squared differences 
are set to the central experimental values: $|\Delta m^2_{32}| = |m_3^2 - m_2^2| =
2.4 \times 10^{-3}$~eV$^2$ and $\Delta m^2_{21} = m_2^2 - m_1^2 = 7.65
\times 10^{-5}$~eV$^2$. The lightest neutrino mass is chosen to be in
the range $\{ 10^{-4}, 10^{-2} \}$~eV. The mixing angles have been
fixed to tribimaximal values. Finally, points satisfying 
$|\left( Y_\nu \right)_{ij}| \leq 1$ are considered. 
For the MSSM, we take $\tan\beta=20$.

\begin{figure}[t!]
\begin{center}
\includegraphics[scale=0.7]{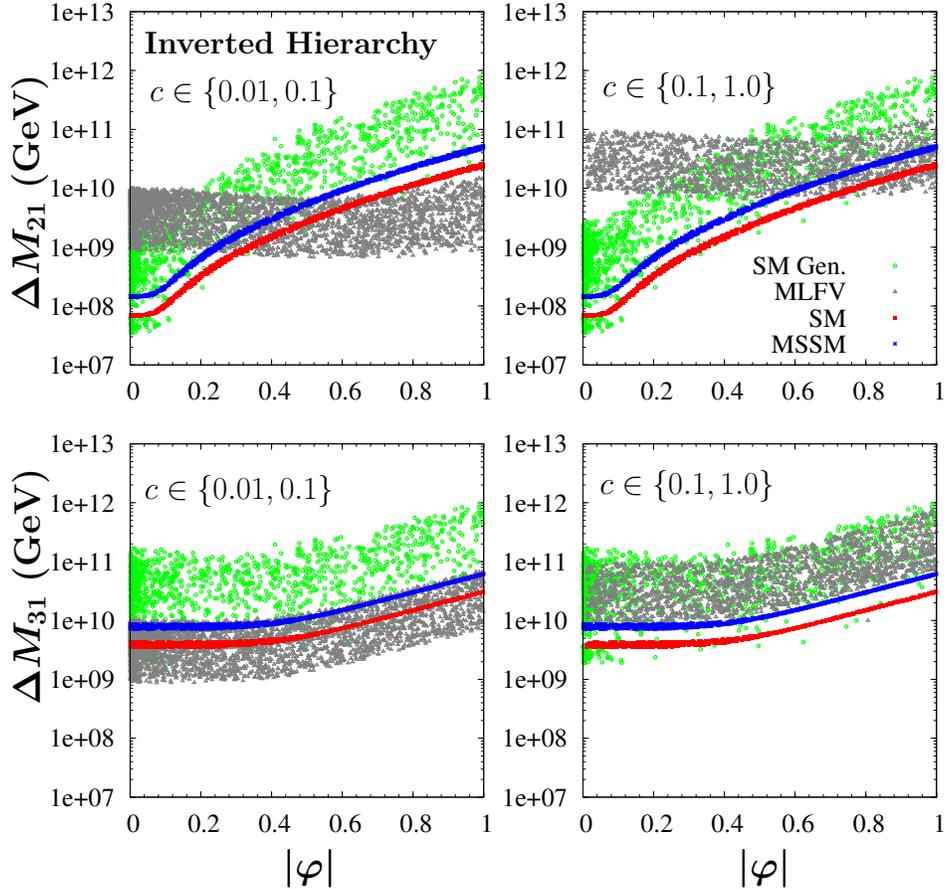}
\caption{Majorana mass splittings as a function 
of $|\varphi|$ for an inverted hierarchy of neutrino masses. 
The green (light gray in black and white) dots show 
`SM Gen.', and dark gray dots are for `MLFV'. The red (lower) 
and blue (upper) dotted lines correspond to SM and MSSM 
respectively.
\label{fig:Type-I-inverted}}
\end{center}
\end{figure}

To illustrate the mass-splitting generated through RG evolution, we 
consider the case of Type-I seesaw and show the results 
when the theory is extended SM, extended MSSM and also 
any generic theory with the same underlying symmetry as 
extended SM (referred to as `SM Gen.'). All these cases 
together are referred to as `Type-I RG'. In case of `SM Gen.', 
we choose the coefficients 
appearing in the evolution of $Y_e$, $Y_\nu$ and $M_\nu$, given in 
Eqs.~(\ref{Ye}), (\ref{Ynu}) and (\ref{RG-Mnu}) respectively, as
\beq
|a_{1,2}|, |b_{1,2}|, q_1, -a_{g_1,g_2}, -b_{g_1,g_2} \in \{0.5,4 \} \; , \qquad 
a_T = b_T = 1 \;, \qquad q_{g_2} = 0 \; .
\eeq
For `Type-I RG', the high scale is chosen to be $\mu_0 = 10^{16}$ GeV, while 
the value of the mass-splitting is evaluated at $\mu = M_R$. 
For the general MLFV scenario \cite{MLFV-leptogenesis-Isidori} 
(referred to as `MLFV'), we consider the case when
\beq
c_{11} = c \qquad {\rm and} \qquad c_{21} = c_{24} = c^2 \; ,
\eeq
with all other $c_n$s set to zero. 
The value of $c$ is varied over a few orders of magnitude, 
$c \in \{ 10^{-2}, 1 \}$, as can be seen in 
Figs.~\ref{fig:Type-I-normal} and \ref{fig:Type-I-inverted}. 
In the `MLFV' scenario, the mass-splitting does not depend on the 
energy scale.

Fig.~\ref{fig:Type-I-normal} shows the plots for normal neutrino 
hierarchy ($\Delta m^2_{32} > 0$), while Fig.~\ref{fig:Type-I-inverted} 
shows that for the inverted case ($\Delta m^2_{32} < 0$).
From the figures one can make the following observations:
\begin{itemize}

\item For both the cases of `MLFV' and `Type-I RG', the nature 
of variation of $\Delta M_{31} = M_3 - M_1$ with $|\varphi|$ is the 
same, for the whole range of $|\varphi| \in \{ 0.001, 1.0\}$, 
with either neutrino mass hierarchy. The generic variation trends are
different for $\Delta M_{21} = M_2 - M_1$. 

\item In case of `Type-I RG', for inverted hierarchy 
$\Delta M_{21}$ varies about two orders of magnitude as $|\varphi|$ is
varied in the range 0.1 -- 1.0. For normal hierarchy, the variation is
small for $|\varphi| \lesssim 0.5$. 
For `MLFV', variation of $\Delta M_{21}$ is quite 
small for inverted hierarchy.

\item There is an overlap of $\Delta M_{21}$ generated 
in `MLFV' for $c \in \{ 0.01, 0.1\}$ with that in `SM Gen.' 
for $|\varphi| > 0.3 (0.15)$ with normal(inverted) hierarchy. 
For higher $c$ values, $c \in \{ 0.1, 1.0 \}$, 
the `MLFV' can resemble the RG effect for the whole 
range of $|\varphi|$ for normal hierarchy, while for inverted 
hierarchy the same is accomplished for $|\varphi| > 0.2$.

\item $\Delta M_{31}$ generated in `MLFV' overlaps that 
in `SM Gen.' for the whole range of $|\varphi|$ with 
both the hierarchies and for all $c \in \{ 0.001, 1.0\}$.

\end{itemize}
The above example shows a consistent treatment of the splitting that
include both the generic splittings from spurion technique and the RG
evolution. 
The result obtained in the case of a general splitting with spurions 
is different from what we get when RG effects are included. 
However, there is an overlap for some region of the parameter 
space.

\begin{figure}[t!]
\begin{center}
\includegraphics[scale=0.7]{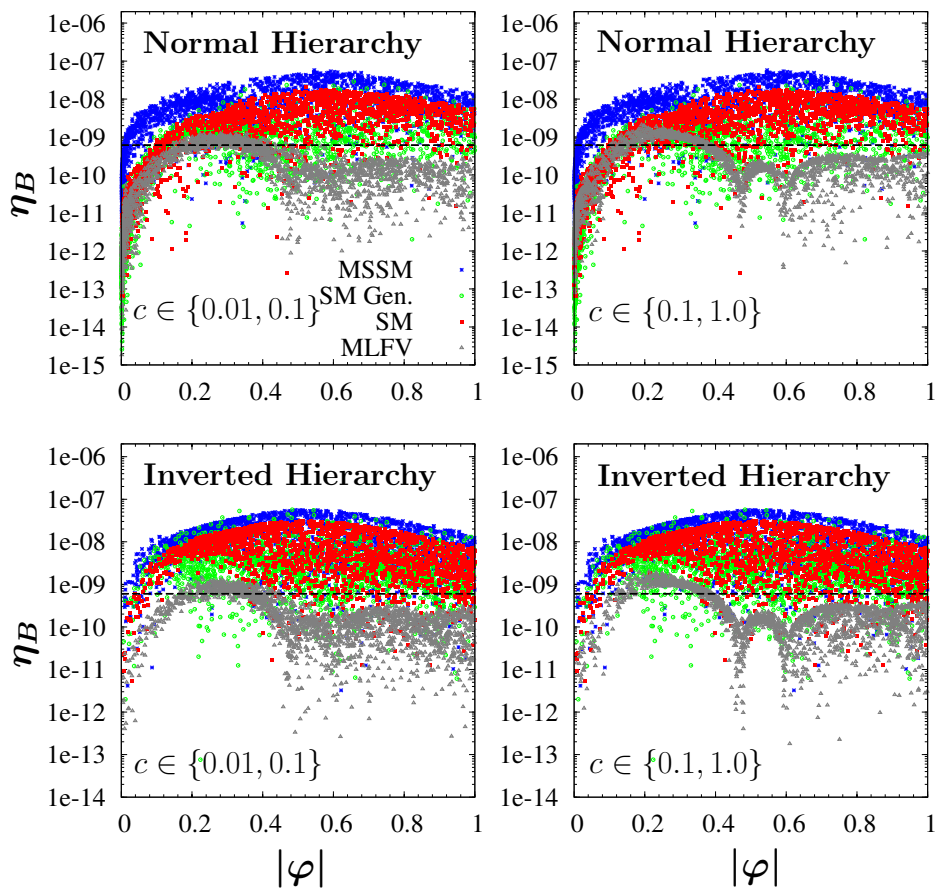}
\caption{Baryon asymmetry of the Universe, $\eta_B$, 
as a function of $|\varphi|$, for `Type-I RG' and `MLFV'.
The black (dashed) horizontal line shows the current 
experimental values of $\eta_B$ at 1$\sigma$. 
\label{fig:etaB}}
\end{center}
\end{figure}
Next, we discuss the effect of including RG evolution on
leptogenesis, and compare it to the result obtained with the generic
splitting~\cite{MLFV-leptogenesis-Isidori}. The baryon asymmetry
$\eta_B$ can be expressed as
\beq
\eta_B = 9.6 \times 10^{-3} \sum_i \epsilon_i d_i \; , 
\eeq
where $d_i$ are the washout factors, and the $\epsilon_i$ 
are the CP asymmetries defined as~ \cite{CPasymmetries,Ellis,Lepto2} 
\beq
\epsilon_i = \frac{\sum_k  \left[ \Gamma(\nu_R^i \to l_k \phi^\ast) - \Gamma(\nu_R^i \to \bar l_k \phi) \right]}
{\sum_k \left[ \Gamma(\nu_R^i \to l_k \phi^\ast) + \Gamma(\nu_R^i \to \bar l_k \phi) \right]} \; .
\eeq
To determine $d_i$, we consider the strong washout regime and 
use the same approximations as in~\cite{MLFV-leptogenesis-Isidori, DiBari}.

Values of $\eta_B$ obtained as a function of $|\varphi|$ 
is shown in Fig.~\ref{fig:etaB}. 
The black (dashed) horizontal line shows the current 
experimental value of the baryon asymmetry~\cite{etaB-PDG}
\beq
\eta_B = (6.23 \pm 0.17) \times 10^{-10} \; ,
\eeq
at 1$\sigma$. It can be seen from Fig.~\ref{fig:etaB} that in case of
generic mass splitting with spurion
techniques~\cite{MLFV-leptogenesis-Isidori} the correct value of
$\eta_B$ can be achieved for $0.1 \lesssim |\varphi| \lesssim 0.4$,
for the given choice of other parameters, with both the neutrino
hierarchies and $c \in \{0.01, 0.1\}$. For other values of
$|\varphi|$, the baryon asymmetry is lower than the current
experimental value. For higher $c$ values, $c \in \{0.1, 1.0 \}$, the
correct $\eta_B$ is obtained for a small region around $|\varphi| \sim
0.1$ and $|\varphi| \sim 0.4$. However, if one considers `Type-I RG',
the correct baryon asymmetry is achieved for the whole $|\varphi|$
range and for both hierarchies. The results obtained in the
two cases are different, with a small overlap in the allowed parameter
space. Hence, while relating the low energy effects with the high
energy phenomena, one must include the complete RG evolution of
parameters, rather than considering a generic mass splitting to mimic
the effect.

%%%%%%%%%%%%%%%%%%%%%%%%%%%%%%%%%%%%%%%%%%%%%%%%%%%%%%%%%
%%%%%%%%%%%%%%%%%%%%%%%%%%%%%%%%%%%%%%%%%%%%%%%%%%%%%%%%%
\section{Conclusion}
\label{sec:conclusion}
%%%%%%%%%%%%%%%%%%%%%%%%%%%%%%%%%%%%%%%%%%%%%%%%%%%%%%%%%
%%%%%%%%%%%%%%%%%%%%%%%%%%%%%%%%%%%%%%%%%%%%%%%%%%%%%%%%%

Neutrino physics provides a window to the physics of very high
scale. In order to learn about high energy physics, one need to use RGEs to connect the low and
high energy scales. In this paper, we study models of MLFV and write the 
RGEs in terms of spurions that capture the whole
effect. It is only the coefficient of each term that varies between models.

Our results serve as a check on the existing calculations. 
For example, we find that both in the SM and MSSM, the difference 
between the right-handed neutrino representations enters only in one term, when 
we consider the evolution of the Yukawa matrix $Y_e$. For the purpose 
of illustration of our results, we have also computed the RGEs of Yukawas 
and masses in case of MSSM Type-III seesaw scenario, for the first time. 
If needed, this spurion analysis method to determine the RG evolution 
can be extended to two loop order, as has been done here, in which 
case we can check where the difference between Type-I and Type-III models 
resides. 
Our results can also be extended to other models. For example, 
in Type-II seesaw and Inverse seesaw, we have more sources of lepton 
flavor breaking. We can include them in the analysis in order to get 
more insight about where the running effects are coming from.

One implication of our results has to do with leptogenesis. Degenerate
right-handed neutrinos cannot give the required baryon asymmetry of the
Universe. Thus, they must be split. The splitting can be accomplished 
in two ways: explicitly with allowed spurion combinations from symmetry 
consideration, as is done in~\cite{MLFV-leptogenesis-Isidori}, or by 
considering RG evolution of different parameters consistently.
We show that the effect of RG running can significantly change the
allowed region of parameter space for successful leptogenesis compared 
to the explicit breaking, and hence should be taken into account.

%%%%%%%%%%%%%%%%%%%%%%%%%%%%%%%%%%%%%%%%%%%%%%%%%%%%%%%%%
%%%%%%%%%%%%%%%%%%%%%%%%%%%%%%%%%%%%%%%%%%%%%%%%%%%%%%%%%
\section*{Acknowledgments}
%%%%%%%%%%%%%%%%%%%%%%%%%%%%%%%%%%%%%%%%%%%%%%%%%%%%%%%%%
%%%%%%%%%%%%%%%%%%%%%%%%%%%%%%%%%%%%%%%%%%%%%%%%%%%%%%%%%

We thank Amol Dighe and Diptimoy Ghosh for useful discussions, 
Joshua Berger for comments on the manuscript. 
This work is supported by the U.S. National Science Foundation through
grant PHY-0757868. 

%%%%%%%%%%%%%%%%%%%%%%%%%%%%%%%%%%%%%%%%%%%%%%%%%%%%%%%%%
%%%%%%%%%%%%%%%%%%%%%%%%%%%%%%%%%%%%%%%%%%%%%%%%%%%%%%%%%
\appendix
%%%%%%%%%%%%%%%%%%%%%%%%%%%%%%%%%%%%%%%%%%%%%%%%%%%%%%%%
%%%%%%%%%%%%%%%%%%%%%%%%%%%%%%%%%%%%%%%%%%%%%%%%%%%%%%%%%

\section{Calculation of RG evolution in MSSM Type-III seesaw}
\label{app:MSSM-typeIII} 
In this section we consider the MSSM extended by the addition 
of three right-handed triplet superfields $\nu$. This is the only
model out of the four we considered where explicit calculation does not
exist in the literature, and thus we present it here.

The Yukawa part of the superpotential is given by
\beqa
{\cal W}_{\rm Yukawa} &=& \left( Y_\nu \right)_{gf} {\nu}^{Cg} {H}_U
{l}^f + \left( Y_e \right)_{gf} {e}^{Cg}
{H}_D {l}^f \nn \\
&+&  \left( Y_U \right)_{gf} {u}^{Cg} 
{H}_U  {Q}^f 
+ \left( Y_D \right)_{gf} {d}^{Cg} 
 {H}_D  {Q}_b^f \; , 
\label{superpot-MSSM}
\eeqa 
where the first line corresponds to the Yukawa interactions for the
lepton superfields, while the second line shows the Yukawa
interactions for the quark superfields. The superfields $e$, $u$ 
and $d$ contain the $\su(2)_L$-singlet charged leptons,
down-type quarks and up-type quarks, while ${l}$ and ${Q}$ contain the
$\su(2)_L$ lepton and quark doublets, respectively. Superpotential
corresponding to the Majorana mass term for triplet neutrino
superfields is
\beqa
{\cal W}_{\rm Maj} &=& \frac{1}{2} {\nu}^{Cg} \left( M_\nu \right)_{gf} {\nu}^{Cf} \; .
\eeqa
${\cal W}_{\rm Maj}$ is important for the seesaw mechanism, but it does not take part 
in the RG evolution of different quantities.

\subsection{Wavefunction renormalization constants}
 
Let us consider a general supersymmetric gauge theory containing 
$N_\Phi$ superfields $\Phi^{(i)}$ that 
transform under the irreducible representations 
${\cal R}_1^{(i)} \times \cdots \times {\cal R}_K^{(i)}$ of the gauge group 
$G_1 \otimes \cdots \otimes G_K$. The renormalizable part 
of the superpotential is given as
\beqa
{\cal W}_{\rm renorm} &=& \frac{1}{6} \sum_{i,j,k=1}^{N_\Phi} 
\lambda_{(ijk)} \Phi^{(i)} \Phi^{(j)} \Phi^{(k)} \; ,
\label{superpot-gen}
\eeqa
where $(ijk)$ implies symmetrization over the indices.
Due to the non-renormalization theorem, the RG evolution equations for 
different operators of the superpotential are governed only by the 
wavefunction renormalization constants for the superfields $\Phi^{(i)}$, 
given as
\beq
Z_{ij} = {\mathbbm I}_{ij} + \delta Z_{ij} \; . 
\eeq
The bare and renormalized superfields, $\Phi^{(i)}_{\mathrm{B}}$ 
and $\Phi^{(i)}$, are then related as
\beq
\Phi^{(i)}_{\mathrm{B}} = \sum_{j=1}^{N_\Phi} 
Z_{ij}^{\frac{1}{2}} \; \Phi^{(j)}\; .
\eeq

Using dimensional regularization {\it via} dimensional reduction, 
the wavefunction renormalization constants, in $d = 4 - \varepsilon$ dimensions, 
at 1-loop are obtained as \cite{Siegel:1979wq, Capper:1979ns}
\beqa
\delta Z_{ij}^{(1)} &=& -\frac{1}{16 \pi^2} \frac{1}{\varepsilon} 
\left[ \sum_{k,l=1}^{N_{\Phi}} \lambda^\ast_{ikl} \lambda_{jkl}
 - 4 \sum_{n=1}^K g_n^2 C_2({\cal R}_n^{(i)}) \delta_{ij}  \right] \; ,
\label{Zij-gen}
\eeqa
where $C_2({\cal R}_n^{(i)})$ is the quadratic Casimir for the representation 
${\cal R}_n^{(i)}$ of the gauge group $G_n$.

Comparing the superpotentials in Eqs.~(\ref{superpot-gen}) and (\ref{superpot-MSSM}), 
and using Eq.~(\ref{Zij-gen}), we get the $1/\varepsilon$ coefficients of the 
wavefunction renormalization constants, for different lepton and Higgs superfields, 
to be
\beqa
&-& \left( 4 \pi\right)^2 \delta Z_l = 
2 Y_e^\dagger Y_e + 2 \left( \sum_a \left( \epsilon G^a \right)^\dagger \left(\epsilon G^a \right) \right) 
Y_\nu^\dagger Y_\nu  - \frac{3}{5} g_1^2 - 3 g_2^2 \; , \label{Zl} \\
&-& \left( 4 \pi\right)^2 \delta Z_{e^C} = 
4 Y_e^\ast Y_e^T - \frac{12}{5} g_1^2  \; , \label{Ze} \\
&-& \left( 4 \pi\right)^2 \delta Z_{\nu^C} = 
2 \; \tr[\left( \epsilon G^a \right)^\dagger \epsilon G^a] \; Y_\nu^\ast Y_\nu^T - 
4 \; C_2({\cal R}_{\su(2)_L}) \; g_2^2  \; , \label{Znu} \\
&-& \left( 4 \pi\right)^2 \delta Z_{H^U} = 
2 \left( \sum_a \left( \epsilon G^a \right)^\ast \left( \epsilon G^a \right)^T \right) \tr[Y_\nu^\dagger Y_\nu]
+ 6 \tr[Y_U^\dagger Y_U] - \frac{3}{5} g_1^2 - 3 g_2^2  \; , \label{ZHu} \\
&-& \left( 4 \pi\right)^2 \delta Z_{H^D} = 
2 \tr[Y_e^\dagger Y_e] + 6 \tr[Y_D^\dagger Y_D] - \frac{3}{5} g_1^2 - 3 g_2^2 \; . \label{ZHd} 
\eeqa 
It must be noted that the wavefunction renormalization constants, 
given in Eqs.~(\ref{Zl}) -- (\ref{Znu}), are in general forms applicable 
to both Type-I and Type-III seesaw when we use appropriate forms of 
$G^a$, as given in Eq.~(\ref{Ga}). 
Thus the quantities, which depend on the transformation properties of the 
right-handed neutrino superfields, are
\beqa
r^\prime &=& \sum_a \left( \epsilon G^a \right)^\ast \left( \epsilon G^a \right)^T 
=  (1,3) \; , \\
\alpha^\prime_1 &=& \sum_a \left( \epsilon G^a \right)^\dagger \left(\epsilon G^a \right) 
= (1,3) \; , \\
\alpha^\prime_2 &=& \tr\left[\left( \epsilon G^a \right)^\dagger \epsilon G^a \right] 
= (2,2)\; . \label{alpha2-prime-app}
\eeqa
Here the numbers in the parenthesis are the values in Type-I and Type-III 
seesaw scenarios, and are universal in the SU(2) space, 
as defined in Eqs.~(\ref{r-prime}) -- (\ref{alpha2-prime}). 
We do not use any summation convention here.
$C_2({\cal R}_{\su(2)_L})$ in Eq.~(\ref{Znu}) is the quadratic Casimir for the superfield 
$\nu$ under $\su(2)_L$ and hence, as given in Eq.~(\ref{casimir}), 
$C_2({\cal R}_{\su(2)_L}) = 0$ for Type-I seesaw, 
and $C_2({\cal R}_{\su(2)_L}) = 2$ for Type-III seesaw. 
In Section~\ref{sec:results} and in the remainder of the appendix we use 
$C_2 \equiv C_2({\cal R}_{\su(2)_L})$.

\subsection{Calculation of RG evolution equations}

Let us now compute the $\beta$-functions. The RG evolution of 
$Y_e$ is given by
\beqa
\mu \frac{d Y_e}{d \mu} &=& -\frac{1}{2} \left( Y_e \delta Z_l + 
Y_e \delta Z_{H_D} + \delta Z_{e^C}^\ast Y_e \right) \; ,
\eeqa
which reduces to
\beqa
\dot Y_e &=& Y_e \left[ 3 Y_e^\dagger Y_e + \alpha_1^\prime Y_\nu^\dagger Y_\nu 
+ \left( \tr[Y_e^\dagger Y_e] + 3 \tr[Y_D^\dagger Y_D] \right) 
- \frac{9}{5} g_1^2 - 3 g_2^2 \right] \nn \\
&=& Y_e \left[ 3 Y_e^\dagger Y_e + \alpha_1^\prime Y_\nu^\dagger Y_\nu 
+ T_D - \frac{9}{5} g_1^2 - 3 g_2^2 \right] \; ,
\eeqa
where
\beq
T_D = \tr[Y_e^\dagger Y_e] + 3 \tr[Y_D^\dagger Y_D] \; .
\eeq 
Similarly, the evolution equation for $Y_\nu$ is given by
\beqa
\dot Y_\nu &=& Y_\nu \left[ Y_e^\dagger Y_e + 
\left( \alpha_1^\prime + \alpha_2^\prime \right) Y_\nu^\dagger Y_\nu 
+ \left( r^\prime \tr[Y_\nu^\dagger Y_\nu] + 3 \tr[Y_U^\dagger Y_U] \right) 
- \frac{3}{5} g_1^2 - \left( 3 + 2 C_2 \right) g_2^2\right] \nn \\
&=& Y_\nu \left[ Y_e^\dagger Y_e + 
\left( \alpha_1^\prime + \alpha_2^\prime \right) Y_\nu^\dagger Y_\nu 
+ T_U - \frac{3}{5} g_1^2 - \left( 3 + 2 C_2 \right) g_2^2\right]
\eeqa
where
\beq
T_U = r^\prime \tr[Y_\nu^\dagger Y_\nu] + 3 \tr[Y_U^\dagger Y_U]  \; .
\eeq  

The evolution equation of the right-handed neutrino mass $M_\nu$ is 
given by
\beqa
\mu \frac{d M_\nu}{d \mu} &=& -\frac{1}{2} \left(  \delta Z_{\nu^C}^T M_\nu 
+ M_\nu \delta Z_{\nu^C} \right) \; ,
\eeqa
which reduces to
\beqa
\dot M_\nu &=& \alpha_2^\prime \left[ \left( Y_\nu Y_\nu^\dagger \right) M_\nu 
+ M_\nu \left( Y_\nu Y_\nu^\dagger \right)^T \right] - 4 C_2 g_2^2 M_\nu \; .
\eeqa

%%%%%%%%%%%%%%%%%%%%%%%%%%%%%%%%%%%%%%%%%%%%%%%%%%%%%%%%%
%%%%%%%%%%%%%%%%%%%%%%%%%%%%%%%%%%%%%%%%%%%%%%%%%%%%%%%%%
\section{RG evolution equations at 2-loop}
\label{app:2loop}
%%%%%%%%%%%%%%%%%%%%%%%%%%%%%%%%%%%%%%%%%%%%%%%%%%%%%%%%%
%%%%%%%%%%%%%%%%%%%%%%%%%%%%%%%%%%%%%%%%%%%%%%%%%%%%%%%%%

In Section~\ref{sec:RG} of the main part of the paper, 
we have considered the first order
contribution of the spurion fields. Here, we study the second order
terms in the RGEs of the Yukawas and the masses
using the same technique.

\subsection{2-loop running of $Y_e$} 

In this section, we consider the evolution of $Y_e$.
The new contributions at 2-loop will consist of five 
spurion fields transforming as $(\bar 3, 3, 1)$ under $\Glf$. 
Any combination of three spurion fields with $(\bar 3, 3, 1)$ 
and two other couplings in the theory transforming 
trivially under $\Glf$ is also a valid term at this order. 
There must also be terms proportional to a single spurion field 
and four other couplings.

Using Table~\ref{tab:trans}, the $\su(3)$-algebra 
\beq
8 \otimes 8 = 27 \oplus 10 \oplus \overline{10} \oplus 8 \oplus 8 \oplus 1 \; , 
\label{su3-algebra-2} 
\eeq
and those given in Eq.~(\ref{su3-algebra}), and the transformation properties 
\beqa
\tr[Y_e^\dagger Y_e Y_e^\dagger Y_e] = (1,1,1) \; , \quad
\tr[Y_\nu^\dagger Y_\nu Y_\nu^\dagger Y_\nu] = (1,1,1) \; , \quad
\tr[Y_e^\dagger Y_e Y_\nu^\dagger Y_\nu] = (1,1,1) \; , 
\label{trace-y4}
\eeqa
we get that 
\beqa
Y_e T_e T_e &=& (\bar 3, 3, 1) \otimes (8,1,1) \otimes (8,1,1) \ni (\bar 3, 3 ,1) \; , \\
Y_e T_e T_\nu &=& (\bar 3, 3, 1) \otimes (8,1,1) \otimes (8,1,1) \ni (\bar 3, 3 ,1) \; , \\
Y_e T_\nu T_e &=& (\bar 3, 3, 1) \otimes (8,1,1) \otimes (8,1,1) \ni (\bar 3, 3 ,1) \; , \\
Y_e T_\nu T_\nu &=& (\bar 3, 3, 1) \otimes (8,1,1) \otimes (8,1,1) \ni (\bar 3, 3 ,1) \; , \\
Y_e \tr[Y_e^\dagger Y_e] T_e &=& (\bar 3, 3, 1) \otimes (1,1,1) \otimes (8,1,1) \ni (\bar 3, 3 ,1) \; , \\
Y_e \tr[Y_\nu^\dagger Y_\nu] T_e &=& (\bar 3, 3, 1) \otimes (1,1,1) \otimes (8,1,1) \ni (\bar 3, 3 ,1) \; , \\
Y_e \tr[Y_e^\dagger Y_e] T_\nu &=& (\bar 3, 3, 1) \otimes (1,1,1) \otimes (8,1,1) \ni (\bar 3, 3 ,1) \; , \\
Y_e \tr[Y_\nu^\dagger Y_\nu] T_\nu &=& (\bar 3, 3, 1) \otimes (1,1,1) \otimes (8,1,1) \ni (\bar 3, 3 ,1) \; , \\
Y_e \tr[Y_e^\dagger Y_e Y_e^\dagger Y_e] &=& (\bar 3, 3, 1) \otimes (1,1,1) = (\bar 3, 3 ,1) \; , \\ 
Y_e \tr[Y_\nu^\dagger Y_\nu Y_\nu^\dagger Y_\nu] &=& (\bar 3, 3, 1) \otimes (1,1,1) = (\bar 3, 3 ,1) \; , \\
{\rm and} \quad 
Y_e \tr[Y_e^\dagger Y_e Y_\nu^\dagger Y_\nu ] &=& (\bar 3, 3, 1) \otimes (1,1,1) = (\bar 3, 3 ,1) \;  
\eeqa
are the only allowed combinations of five spurion fields 
that can appear on the RHS of $\dot Y_e$ at second order.
Hence we can write the most general form of the 
second order contributions to $\dot Y_e$ as
\begin{figure}[t!]
\begin{center}
\includegraphics[scale=0.9]{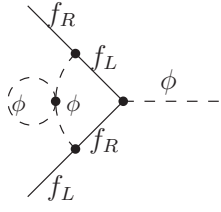}
\caption{Example of diagram contributing terms proportional to 
$d_{12}^\lambda$, $d_{13}^\lambda$.
\label{fig:lambda}}
\end{center}
\end{figure}
\beqa
\left.(4\pi)^2 \; \dot Y_e \right\arrowvert_{\text{2-loop}} &\sim& 
Y_e \left(\widetilde d_1 T_e T_e +\widetilde d_2 T_e T_\nu 
+ \widetilde d_3 T_\nu T_e + \widetilde d_4 T_\nu T_\nu \right) \nn \\
&+& Y_e \left( \widetilde d_5 \tr[Y_e^\dagger Y_e] T_e 
+ \widetilde d_6 \tr[Y_\nu^\dagger Y_\nu] T_e
+ \widetilde d_7 \tr[Y_e^\dagger Y_e] T_\nu
+ \widetilde d_8 \tr[Y_\nu^\dagger Y_\nu] T_\nu \right) \nn \\
&+& Y_e \left(\widetilde d_9 \tr[Y_e^\dagger Y_e Y_e^\dagger Y_e] 
+\widetilde d_{10} \tr[Y_\nu^\dagger Y_\nu Y_\nu^\dagger Y_\nu] 
+\widetilde d_{11} \tr[Y_e^\dagger Y_e Y_\nu^\dagger Y_\nu ]\right) \nn \\
&+& Y_e \left( \widetilde d_{12} T_e + \widetilde d_{13} T_\nu + 
\widetilde d_{14} \tr[Y_e^\dagger Y_e] +\widetilde d_{15}\tr[Y_\nu^\dagger Y_\nu] \right) 
+ \widetilde d_{16} Y_e \; .
\label{Ye-2loop-1}
\eeqa 
The extra factor of $(4\pi)^2$ is there since we are considering 
2-loop contributions.
We rewrite Eq.~(\ref{Ye-2loop-1}), using the definitions of $T_e$, $T_\nu$ 
from Table~\ref{tab:trans}, as
\beqa
\left. (4\pi)^2 \; \dot Y_e \right\arrowvert_{\text{2-loop}} &=& 
Y_e \left( d_1 Y_e^\dagger Y_e Y_e^\dagger Y_e + d_2 Y_e^\dagger Y_e Y_\nu^\dagger Y_\nu 
+ d_3 Y_\nu^\dagger Y_\nu Y_e^\dagger Y_e + d_4 Y_\nu^\dagger Y_\nu Y_\nu^\dagger Y_\nu \right) \nn \\
&+& Y_e \left( d_5 Y_e^\dagger Y_e \tr[Y_e^\dagger Y_e] + d_6 Y_e^\dagger Y_e \tr[Y_\nu^\dagger Y_\nu] 
+ d_7 Y_\nu^\dagger Y_\nu \tr[Y_e^\dagger Y_e] + d_8 Y_\nu^\dagger Y_\nu \tr[Y_\nu^\dagger Y_\nu]\right) \nn \\
&+& Y_e \left( d_9 \tr[Y_e^\dagger Y_e Y_e^\dagger Y_e] 
+ d_{10} \tr[Y_\nu^\dagger Y_\nu Y_\nu^\dagger Y_\nu] 
+ d_{11} \tr[Y_e^\dagger Y_e Y_\nu^\dagger Y_\nu ] \right) \nn \\
&+& Y_e \left( d_{12} Y_e^\dagger Y_e + d_{13} Y_\nu^\dagger Y_\nu + d_{14} \tr[Y_e^\dagger Y_e] 
+ d_{15}\tr[Y_\nu^\dagger Y_\nu] \right) + d_{16} Y_e \; ,
\eeqa
where $d_1$, $\cdots$, $d_{11}$ are expected to be ${\cal O}(1)$ numbers. 
We have not written the terms of the form
$\tr[Y_i^\dagger Y_i] \cdot \tr[Y_j^\dagger Y_j]$, $i,j \in \{e,\nu\}$, since
such terms cannot be generated at 2-loop order.
Each of $d_{12}$, $d_{13}$ is expected to be a linear function of
$g_i^2$, $\lambda$, $\tr[Y_U^\dagger Y_U]$, $\tr[Y_D^\dagger Y_D]$ and
can be written, in general, as
\beq
d_i = d_i^{g_1} g_1^2 + d_i^{g_2} g_2^2 + d_i^\lambda \lambda + d_i^U
\tr[Y_U^\dagger Y_U] + d_i^D \tr[Y_D^\dagger Y_D] \qquad ( i \in \{12,
13\} ) \; .
\label{d-12-13}
\eeq
Unlike the 1-loop case, Higgs self-coupling can appear at 2-loop order
via diagrams like the one shown in Fig.~\ref{fig:lambda}. Since the
leptons are singlets under $\su(3)_C$, $g_3^2$ cannot be present in
$d_{12}$ and $d_{13}$. As before, $d_{12}^x$, $d_{13}^x$ are expected
to be of ${\cal O}(1)$.
$d_{14}$, $d_{15}$ must originate from a diagram containing 
complete lepton loop in Higgs self-energy correction and hence 
cannot contain $\lambda$ or $g_3^2$. Hence we write
\beq
d_i = d_i^{g_1} g_1^2 + d_i^{g_2} g_2^2  \qquad ( i \in \{14, 15\} ) \; , 
\label{d-14-15}
\eeq
where $d_{14}^x$, $d_{15}^x$ are to be of ${\cal O}(1)$. 
$\tr[Y_U^\dagger Y_U]$ or $\tr[Y_D^\dagger Y_D]$ cannot be present in 
$d_{14}$ and $d_{15}$.

\begin{figure}[t!]
\begin{center}
\includegraphics[scale=0.9]{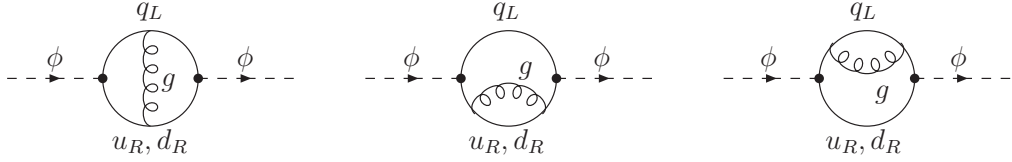}
\caption{Example of diagrams at 2-loop 
with gluon contributions, leading to terms proportional to 
$g_3^2 \tr[Y_U^\dagger Y_U]$ and $g_3^2 \tr[Y_D^\dagger Y_D]$ in $d_{16}$. 
\label{fig:g3}}
\end{center}
\end{figure}
\begin{figure}[t!]
\begin{center}
\includegraphics[scale=0.9]{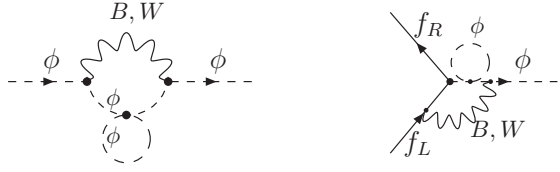}
\caption{Example of diagrams at 2-loop 
giving rise to $\lambda g_1^2$,$\lambda g_2^2$ terms in $d_{16}$.
\label{fig:lambda-gi}}
\end{center}
\end{figure}

Let us now consider the quantity $d_{16}$, which is independent 
of the spurion fields and must be a function 
linear in $\tr[Y_U^\dagger Y_U Y_U^\dagger Y_U]$, $\tr[Y_D^\dagger Y_D Y_D^\dagger Y_D]$ and 
quadratic in $g_i^2$, $\tr[Y_U^\dagger Y_U]$, $\tr[Y_D^\dagger Y_D]$ and $\lambda$.
In its most general form, it can be expressed as
\beqa
d_{16} &=& d_{16}^{UU} \tr[Y_U^\dagger Y_U Y_U^\dagger Y_U]
+ d_{16}^{DD} \tr[Y_D^\dagger Y_D Y_D^\dagger Y_D] 
+ d_{16}^{UD} \tr[Y_U^\dagger Y_U Y_D^\dagger Y_D]  \nn \\
&+& \left( d_{16}^{g_1 U} g_1^2 + d_{16}^{g_2 U} g_2^2 + d_{16}^{g_3 U} g_3^2 \right) \tr[Y_U^\dagger Y_U] + 
\left( d_{16}^{g_1 D} g_1^2 + d_{16}^{g_2 D} g_2^2 + d_{16}^{g_3 D} g_3^2 \right) \tr[Y_D^\dagger Y_D]  \nn \\
&+& \left( d_{16}^{g_1 \lambda} g_1^2 + d_{16}^{g_2 \lambda} g_2^2 \right) \lambda
+ d_{16}^{g_1} g_1^4 + d_{16}^{g_2} g_2^4 + d_{16}^{g_1 g_2} g_1^2 g_2^2  \; ,
\label{d-16}
\eeqa
where all the coefficients $d_{16}^x$ are expected to be ${\cal O}(1)$ numbers. 
Unlike the case of first order evolution equation, 
here $g_3^2$ can appear at 2-loop since quarks have color charges.
For example, diagrams shown in Fig.~\ref{fig:g3} 
will contribute terms proportional to 
$g_3^2 \tr[Y_U^\dagger Y_U]$ and $g_3^2 \tr[Y_D^\dagger Y_D]$.
However, terms proportional to $g_3^4$ cannot be present. 
As can be checked, here we cannot have terms proportional to 
$\lambda \tr[Y_U^\dagger Y_U]$ or $\lambda \tr[Y_D^\dagger Y_D]$, 
while terms containing $\lambda g_1^2$,$\lambda g_2^2$ can 
contribute. Examples of diagrams giving rise to such terms 
are shown in Fig.~\ref{fig:lambda-gi}.
There cannot exist any term proportional to $\lambda g_3^2$ or 
$\lambda^2$ in this case.

Having written the most general form of second order contributions to
$\dot Y_e$, we consider the fact that $\tr[Y_i^\dagger Y_i]$ ($i \in
\{e, \nu, U,D\}$) can only come from a complete fermionic loop in
the Higgs self-energy correction, as already stated in
Section~\ref{sec:RG-Ye-1} and shown in Fig.~\ref{Higgs-self}. Hence, we
can write the ratios as
\beqa
d_5 : d_6 : d_{12}^U : d_{12}^D = 1: r: 3:3 \; , \nn \\
d_7 : d_8 : d_{13}^U : d_{13}^D = 1: r: 3:3 \; ,
\eeqa
where $r$ for Type-I and Type-III seesaw is defined in Eq.~(\ref{r-SM}) 
for SM and in Eq.~(\ref{r-prime}) for MSSM. Hence, we can write
\beqa
d_5 \tr[Y_e^\dagger Y_e] + d_6 \tr[Y_\nu^\dagger Y_\nu] + 
d_{12}^U \tr[Y_U^\dagger Y_U] + d_{12}^D \tr[Y_D^\dagger Y_D] \to d_{12}^T T \; , \nn \\
d_7 \tr[Y_e^\dagger Y_e] + d_8 \tr[Y_\nu^\dagger Y_\nu] + 
d_{13}^U \tr[Y_U^\dagger Y_U] + d_{13}^D \tr[Y_D^\dagger Y_D] \to d_{13}^T T \; , \nn 
\eeqa
where $T$ is defined in Eq.~(\ref{T}) and $d_{13}^T$, $d_{14}^T$ are 
expected to be of ${\cal O}(1)$.
Thus the most general form of $\dot Y_e$ becomes
{\small{
\beqa
\left. (4\pi)^2 \,\dot Y_e \right\arrowvert_{\text{2-loop}} &=& 
Y_e \left( d_1 Y_e^\dagger Y_e Y_e^\dagger Y_e + d_2 Y_e^\dagger Y_e Y_\nu^\dagger Y_\nu 
+ d_3 Y_\nu^\dagger Y_\nu Y_e^\dagger Y_e + 
d_4 Y_\nu^\dagger Y_\nu Y_\nu^\dagger Y_\nu \right) \nn \\
&+& Y_e \left( d_9 \tr[Y_e^\dagger Y_e Y_e^\dagger Y_e] 
+ d_{10} \tr[Y_\nu^\dagger Y_\nu Y_\nu^\dagger Y_\nu] 
+ d_{11} \tr[Y_e^\dagger Y_e Y_\nu^\dagger Y_\nu ] \right) \nn \\
&+& Y_e \left( d_{12}^{g_1} g_1^2 + d_{12}^{g_2} g_2^2 +  d_{12}^\lambda \lambda + d_{12}^T T \right) Y_e^\dagger Y_e + 
Y_e \left(d_{13}^{g_1} g_1^2 + d_{14}^{g_2} g_2^2 +  d_{13}^\lambda \lambda + d_{13}^T T \right) Y_\nu^\dagger Y_\nu \nn \\
&+& Y_e \left( d_{14} \tr[Y_e^\dagger Y_e] + d_{15}\tr[Y_\nu^\dagger Y_\nu] \right) + d_{16} Y_e \; .
\label{Ye-2loop}
\eeqa
}}

%%%%%%%%%%%%%%%%%%%%%%%%%%%%%%%%%%%%%%%%%%%%%%%%%%%%%%%%%
%%%%%%%%%%%%%%%%%%%%%%%%%%%%%%%%%%%%%%%%%%%%%%%%%%%%%%%%%
\subsection{2-loop running of $Y_\nu$} 
%%%%%%%%%%%%%%%%%%%%%%%%%%%%%%%%%%%%%%%%%%%%%%%%%%%%%%%%%
%%%%%%%%%%%%%%%%%%%%%%%%%%%%%%%%%%%%%%%%%%%%%%%%%%%%%%%%%

Let us now consider the second order terms 
arising in the RGE of $Y_\nu$. 
Considering Table~\ref{tab:trans}, the transformation 
rules in Eqs.~(\ref{su3-algebra}, \ref{su3-algebra-2}), 
and the transformation properties in Eq.~(\ref{trace-y4}), 
we get that
\beqa
Y_\nu T_e T_e &=& (\bar 3, 1, 3) \otimes (8,1,1) \otimes (8,1,1) \ni (\bar 3, 1, 3) \; , \\
Y_\nu T_e T_\nu &=& (\bar 3, 1, 3) \otimes (8,1,1) \otimes (8,1,1) \ni (\bar 3,1, 3) \; , \\
Y_\nu T_\nu T_e &=& (\bar 3, 1, 3) \otimes (8,1,1) \otimes (8,1,1) \ni (\bar 3, 1, 3) \; , \\
Y_\nu T_\nu T_\nu &=& (\bar 3, 1, 3) \otimes (8,1,1) \otimes (8,1,1) \ni (\bar 3,1, 3) \; , \\
Y_\nu \tr[Y_e^\dagger Y_e] T_e &=& (\bar 3, 1, 3) \otimes (1,1,1) \otimes (8,1,1) \ni (\bar 3, 1, 3) \; , \\
Y_\nu \tr[Y_\nu^\dagger Y_\nu] T_e &=& (\bar 3, 1, 3) \otimes (1,1,1) \otimes (8,1,1) \ni (\bar 3, 1, 3) \; , \\
Y_\nu \tr[Y_e^\dagger Y_e] T_\nu &=& (\bar 3, 1, 3) \otimes (1,1,1) \otimes (8,1,1) \ni (\bar 3, 1, 3) \; , \\
Y_\nu \tr[Y_\nu^\dagger Y_\nu] T_\nu &=& (\bar 3, 1, 3) \otimes (1,1,1) \otimes (8,1,1) \ni (\bar 3, 1, 3) \; , \\
Y_\nu \tr[Y_e^\dagger Y_e Y_e^\dagger Y_e] &=& (\bar 3, 1, 3) \otimes (1,1,1) = (\bar 3, 1, 3) \; , \\
Y_\nu \tr[Y_\nu^\dagger Y_\nu Y_\nu^\dagger Y_\nu] &=& (\bar 3, 1, 3) \otimes (1,1,1) = (\bar 3, 1, 3)\; , \\ 
{\rm and} \quad 
Y_\nu \tr[Y_e^\dagger Y_e Y_\nu^\dagger Y_\nu] &=& (\bar 3, 1, 3) \otimes (1,1,1) = (\bar 3, 1, 3)\;  
\eeqa
are the only allowed combinations of five spurion fields 
that can appear on the RHS of $\dot Y_\nu$ at second order.
Hence, similar to $\dot Y_e$, we can write the most general form of the 
second order contributions to $\dot Y_\nu$ as
\beqa
\left. (4\pi)^2 \; \dot Y_\nu \right\arrowvert_{\text{2-loop}} &\sim& 
Y_\nu \left(\widetilde f_1 T_e T_e +\widetilde f_2 T_e T_\nu 
+ \widetilde f_3 T_\nu T_e + \widetilde f_4 T_\nu T_\nu \right) \nn \\
&+& Y_\nu \left(\widetilde f_5 \tr[Y_e^\dagger Y_e] T_e +
\widetilde f_6 \tr[Y_\nu^\dagger Y_\nu] T_e +
\widetilde f_7 \tr[Y_e^\dagger Y_e] T_\nu +
\widetilde f_8 \tr[Y_\nu^\dagger Y_\nu] T_\nu \right) \nn \\
&+& Y_\nu \left(\widetilde f_9 \tr[Y_e^\dagger Y_e Y_e^\dagger Y_e] 
+\widetilde f_{10} \tr[Y_\nu^\dagger Y_\nu Y_\nu^\dagger Y_\nu] 
+\widetilde f_{11} \tr[Y_e^\dagger Y_e Y_\nu^\dagger Y_\nu ] \right) \nn \\
&+& Y_\nu \left( \widetilde f_{12} T_e + \widetilde f_{13} T_\nu 
+ \widetilde f_{14} \tr[Y_e^\dagger Y_e] +\widetilde f_{15}\tr[Y_\nu^\dagger Y_\nu] \right) 
+ \widetilde f_{16} Y_e \; .
\eeqa 
The above equation can be written in a simple form 
using the definitions of $T_e$, $T_\nu$ from Table~\ref{tab:trans} and 
the ratio of the coefficient of the traces, as done 
in case of $\dot Y_e$, to give
{\small{
\beqa
\left. (4\pi)^2 \; \dot Y_\nu \right\arrowvert_{\text{2-loop}} &=& 
Y_\nu \left( f_1 Y_e^\dagger Y_e Y_e^\dagger Y_e + f_2 Y_e^\dagger Y_e Y_\nu^\dagger Y_\nu 
+ f_3 Y_\nu^\dagger Y_\nu Y_e^\dagger Y_e + f_4 Y_\nu^\dagger Y_\nu Y_\nu^\dagger Y_\nu \right) \nn \\
&+& Y_\nu \left( f_9 \tr[Y_e^\dagger Y_e Y_e^\dagger Y_e] 
+ f_{10} \tr[Y_\nu^\dagger Y_\nu Y_\nu^\dagger Y_\nu] 
+ f_{11} \tr[Y_e^\dagger Y_e Y_\nu^\dagger Y_\nu ] \right) \nn \\
&+& Y_\nu \left( f_{12}^{g_1} g_1^2 + f_{13}^{g_2} g_2^2 + f_{12}^\lambda \lambda + f_{12}^T T \right) Y_e^\dagger Y_e + 
Y_\nu \left(f_{13}^{g_1} g_1^2 + f_{13}^{g_2} g_2^2 + f_{13}^\lambda \lambda + f_{13}^T T \right) Y_\nu^\dagger Y_\nu \nn \\
&+& Y_\nu \left( f_{14} \tr[Y_e^\dagger Y_e] + f_{15} \tr[Y_\nu^\dagger Y_\nu] \right) + f_{16} Y_\nu \; ,
\label{Ynu-2loop}
\eeqa
}}
with $T$ defined in Eq.~(\ref{T}). 
Here, $f_{1,\cdots,4}$, $f_{9,\cdots,11}$, $f_{12}^x$ and $f_{13}^x$ 
are expected to be ${\cal O}(1)$ numbers.
$f_{i}$ ($i$=14,15,16) will have similar forms as $d_i$ ($i$=14,15,16), 
as given in Eqs.~(\ref{d-14-15}) and (\ref{d-16}) respectively, 
with all $f_i^x$ ($i$=14,15,16) being ${\cal O}(1)$ quantities.
As before, we have not written the terms of the form $\tr[Y_i^\dagger Y_i] \cdot \tr[Y_j^\dagger Y_j]$, 
$i,j \in \{e,\nu, U,D \}$, since such terms cannot be generated at 2-loop.

%%%%%%%%%%%%%%%%%%%%%%%%%%%%%%%%%%%%%%%%%%%%%%%%%%%%%%%%%
%%%%%%%%%%%%%%%%%%%%%%%%%%%%%%%%%%%%%%%%%%%%%%%%%%%%%%%%%
\subsection{2-loop running of $M_\nu$} 
%%%%%%%%%%%%%%%%%%%%%%%%%%%%%%%%%%%%%%%%%%%%%%%%%%%%%%%%%
%%%%%%%%%%%%%%%%%%%%%%%%%%%%%%%%%%%%%%%%%%%%%%%%%%%%%%%%%

Next, we discuss the second order contribution to $\dot M_\nu$.
Using Table~\ref{tab:trans} and the $\su(3)$ algebra given in 
Eqs.~(\ref{su3-mnu}) and (\ref{su3-algebra-2}), we obtain that
\beqa
M_\nu T_\nu^\prime T_\nu^\prime &=& (1,1,6) \otimes 
(1, 1, 8 ) \otimes (1, 1, 8) \ni (1,1,6) \; , \\
M_\nu \left( Y_\nu Y_e^\dagger Y_e Y_\nu^\dagger \right) &=& (1,1,6) \otimes 
(\bar 3, 1,3) \otimes ( 8 \oplus 1, 1, 1) \otimes (3, 1, \bar 3) \ni (1,1,6) \; , 
\label{new-term} \\
T_\nu^{\prime T} M_\nu T_\nu^\prime 
&=& (1, 1, 8) \otimes (1,1,6) \otimes (1, 1, 8) \ni (1,1,6) \; , \\
M_\nu \tr[Y_e^\dagger Y_e] T_\nu^\prime &=& (1,1,6) \otimes 
(1,1,1) \otimes (1, 1, 8) \ni (1,1,6) \; , \\
{\rm and} \quad 
M_\nu \tr[Y_\nu^\dagger Y_\nu] T_\nu^\prime &=& (1,1,6) \otimes 
(1,1,1) \otimes (1, 1, 8) \ni (1,1,6) 
\eeqa
are the only combinations of five spurion fields that can 
contribute to $\dot M_\nu$. 
The term in Eq.~(\ref{new-term}), not present in Table~\ref{tab:trans}, 
is an allowed combination at second order. 
Here we have considered the fact that 
$M_\nu$ couples only to the right-handed neutrinos and hence 
$\dot M_\nu$ cannot contain trace of four spurions at second order. 
Apart from the above terms, there will also be terms 
with three spurions and two other 
couplings in the theory, transforming trivially under $\Glf$. 
Terms containing one spurion and four other couplings are 
also allowed at this order. However, $M_\nu$ being coupled to 
right-handed neutrinos alone, $\dot M_\nu$ will not contain 
terms proportional to trace of four $Y_{U,D}$ and also no $g_1^4$ or $\lambda^2$. 
If the right-handed neutrinos are singlets under the gauge group, 
as is the case for Type-I seesaw, they will not have any 
$\su(2)_L$ or $\su(3)_C$ charges and 
hence terms proportional to $g_2^4$, $g_3^4$ be absent. However, 
for Type-III seesaw scenario these are triplet under 
$\su(2)_L$ and hence $g_2^4$ contribution is expected to be there.

Finally, symmetrizing over the O(3)$_{\nu_R}$ indices, 
the most general form of the 2-loop contribution to $\dot M_\nu$ 
can be written as
{\small{
\beqa
\left. (4\pi)^2 \; \dot M_\nu \right\arrowvert_{\text {2-loop}} &=&
\frac{\widetilde h_1}{2} \left( M_\nu \left(T_\nu^\prime T_\nu^\prime \right) 
+ \left(T_\nu^\prime T_\nu^\prime \right)^T M_\nu \right) + 
\frac{h_2}{2} \left( M_\nu \left( Y_\nu Y_e^\dagger Y_e Y_\nu^\dagger \right)
+ \left( Y_\nu Y_e^\dagger Y_e Y_\nu^\dagger \right)^T M_\nu  \right) \nn \\
&+& \widetilde h_3 T_\nu^{\prime T} M_\nu T_\nu^\prime + 
\frac{1}{2} \left( \widetilde h_4^\prime \tr[Y_e^\dagger Y_e] + \widetilde h_4^{\prime \prime} \tr[Y_\nu^\dagger Y_\nu] \right) 
\left( M_\nu T_\nu^\prime + T_\nu^{\prime T} M_\nu \right) \nn \\
&+& \frac{\widetilde h_4}{2} \left( M_\nu T_\nu^\prime + T_\nu^{\prime T} M_\nu \right) + \widetilde h_5 g_2^4 M_\nu \; .
\label{Mnu-2loop-1}
\eeqa}}
Eq.~(\ref{Mnu-2loop-1}) can be simplified using the definition of $T_\nu^\prime$ 
and the fact that terms proportional to $\tr[Y_i^\dagger Y_i]~(i \in \{ e,\nu, U, D \})$ 
appear only in the combination $T$, defined in Eq.~(\ref{T}), to get 
\beqa
\left. (4\pi)^2 \;\dot M_\nu \right\arrowvert_{\text {2-loop}} &=& 
h_1 \left( M_\nu \left( Y_\nu Y_\nu^\dagger Y_\nu Y_\nu^\dagger \right)
+ \left( Y_\nu Y_\nu^\dagger Y_\nu Y_\nu^\dagger \right)^T M_\nu \right) \nn \\
&+& h_2 \left( M_\nu \left( Y_\nu Y_e^\dagger Y_e Y_\nu^\dagger \right)
+ \left( Y_\nu Y_e^\dagger Y_e Y_\nu^\dagger \right)^T M_\nu  \right) + 
h_3 \left( Y_\nu Y_\nu^\dagger \right)^T M_\nu \left( Y_\nu Y_\nu^\dagger \right) \nn \\
&+& h_4 \left( M_\nu \left( Y_\nu Y_\nu^\dagger \right) 
+ \left( Y_\nu Y_\nu^\dagger \right)^T M_\nu \right) + h_5 g_2^4 M_\nu \; ,
\label{Mnu-2loop}
\eeqa
where $h_1$, $h_2$, $h_3$ and $h_5$ are expected to be ${\cal O}(1)$ 
numbers in general. For Type-I seesaw, $h_5=0$. 
In writing Eq.~(\ref{Mnu-2loop}), we have considered the fact that terms with 
$\tr[Y_i^\dagger Y_i] \cdot \tr[Y_j^\dagger Y_j] (i,j \in \{ e,\nu \})$ 
cannot be present at 2-loop. $h_4$ can in general be a linear function of 
$g_1^2$, $g_2^2$, $\lambda$ and $T$ and be given by
\beq
h_4 = h_4^{g_1} g_1^2 + h_4^{g_2} g_2^2 + h_4^\lambda \lambda + h_4^T T \; ,
\label{h4}
\eeq
where all $h_4^x$ must be of ${\cal O}(1)$. 
In writing Eq.~(\ref{Mnu-2loop}) we have used the symmetry 
property of $M_\nu$: $M_\nu^T = M_\nu$. 
Leptons and Higgs, being singlets under $\su(3)_C$, 
$h_4$ will not involve $g_3^2$.

As before, we expect Eq.~(\ref{Mnu-2loop}) to give the 
right-handed projection of $\dot M_\nu$ only. 
% This is because of the fact that we have considered 
% the right-handed Majorana mass matrix to be real (and universal). 
The most general form of $\dot M_\nu$ will be given by
{\footnotesize{
\beqa
 (4\pi)^2 \left.\dot M_\nu \right\arrowvert_{\text {\tiny{2-loop}}} &=& 
h_1 \left[ \left( M_\nu \left( Y_\nu Y_\nu^\dagger Y_\nu Y_\nu^\dagger \right)
+ \left( Y_\nu Y_\nu^\dagger Y_\nu Y_\nu^\dagger \right)^T M_\nu \right) P_R 
+ \left( M_\nu \left( Y_\nu Y_\nu^\dagger Y_\nu Y_\nu^\dagger \right)^T
+ \left( Y_\nu Y_\nu^\dagger Y_\nu Y_\nu^\dagger \right) M_\nu \right) P_L \right]
 \nn \\
&+& h_2 \left[ \left( M_\nu \left( Y_\nu Y_e^\dagger Y_e Y_\nu^\dagger \right)
+ \left( Y_\nu Y_e^\dagger Y_e Y_\nu^\dagger \right)^T M_\nu  \right)  P_R + 
\left( M_\nu \left( Y_\nu Y_e^\dagger Y_e Y_\nu^\dagger \right)^T
+ \left( Y_\nu Y_e^\dagger Y_e Y_\nu^\dagger \right) M_\nu  \right)  P_L \right] \nn \\
&+& h_3 \left[ \left( Y_\nu Y_\nu^\dagger \right)^T M_\nu \left( Y_\nu Y_\nu^\dagger \right) P_R 
+ \left( Y_\nu Y_\nu^\dagger \right) M_\nu \left( Y_\nu Y_\nu^\dagger \right)^T P_L \right] \nn \\
&+& h_4 \left[ \left( M_\nu \left( Y_\nu Y_\nu^\dagger \right) 
+ \left( Y_\nu Y_\nu^\dagger \right)^T M_\nu \right) P_R + 
\left( M_\nu \left( Y_\nu Y_\nu^\dagger \right)^T + 
\left( Y_\nu Y_\nu^\dagger \right) M_\nu \right) P_L \right]+ h_5 g_2^4 M_\nu \; .
\eeqa
}}

%%%%%%%%%%%%%%%%%%%%%%%%%%%%%%%%%%%%%%%%%%%%%%%%%%%%%%%%%
%%%%%%%%%%%%%%%%%%%%%%%%%%%%%%%%%%%%%%%%%%%%%%%%%%%%%%%%%
\subsection{2-loop running of the left-handed mass $m_\nu$ at $\mu < M_R$} 
%%%%%%%%%%%%%%%%%%%%%%%%%%%%%%%%%%%%%%%%%%%%%%%%%%%%%%%%%
%%%%%%%%%%%%%%%%%%%%%%%%%%%%%%%%%%%%%%%%%%%%%%%%%%%%%%%%%

At the energy scale $\mu < M_R$, the flavor symmetry group is 
$\Glf^\prime$ and $Y_e(\bar 3, 3)$, $m_\nu (6,1)$ are 
the only spurions in the theory. 
Let us first consider the running of $Y_e$ at this scale 
which can be obtained from Eq.~(\ref{Ye-2loop}) simply 
by setting the coefficients of terms containing $Y_\nu$ to zero 
and we have
\beqa
\left. (4\pi)^2 \; \dot Y_e \right\arrowvert_{\text{2-loop}} &=& 
Y_e \left( d_1 Y_e^\dagger Y_e Y_e^\dagger Y_e + 
d_9 \tr[Y_e^\dagger Y_e Y_e^\dagger Y_e] + 
\left( d_{12}^{g_1} g_1^2 + d_{12}^{g_2} g_2^2 +  d_{12}^\lambda \lambda + d_{12}^T T \right) Y_e^\dagger Y_e\right) \nn \\
&+&  Y_e \left( d^{g_1}_{14} g_1^2 + d^{g_2}_{14} g_2^2 \right) \tr[Y_e^\dagger Y_e]  + d_{16} Y_e \; ,
\eeqa
where $d_{16}$ is given by Eq.~(\ref{d-16}). $d_1$, $d_9$, $d_{12}^x$ and 
$d_{14}^x$ are expected to be ${\cal O}(1)$ numbers.

Now we consider the running of the left-handed mass $m_\nu$.
Using Table~\ref{tab:trans}, 
the transformation rules in Eq.~(\ref{trace-y4}) and the 
$\su(3)$ algebra given in Eqs.~(\ref{su3-mnu}, \ref{su3-algebra-2}) 
we get the second order contributions to $\dot m_\nu$ 
to contain the following combinations of five spurions:
\beqa
m_\nu T_e T_e &=& (6,1) \otimes (8,1) \otimes (8,1) \ni (6,1) \; , \\
T_e^T m_\nu T_e &=& (8,1) \otimes (6,1) \otimes (8,1)\ni (6,1) \; , \\
m_\nu \tr[Y_e^\dagger Y_e] T_e &=& (6,1) \otimes (1,1) \otimes (8,1) \ni (6,1) \; , \\
{\rm and} \quad m_\nu \tr[Y_e^\dagger Y_e Y_e^\dagger Y_e] &=& (6,1) \otimes (1,1) = (6,1) \; ,  
\eeqa
where the terms proportional to $\tr[Y_e^\dagger Y_e] \cdot \tr[Y_e^\dagger Y_e]$ 
are to be removed since such terms cannot arise at 2-loop. Finally, 
symmetrizing over the $\su(3)_{l_L}$ indices, we write down the 
most general form of $\dot m_\nu$ at second order as
\beqa
\left. (4\pi)^2 \; \dot m_\nu \right\arrowvert_{\text{2-loop}} &=& 
r_1 \left( m_\nu \left(Y_e^\dagger Y_e Y_e^\dagger Y_e\right) + 
\left(Y_e^\dagger Y_e Y_e^\dagger Y_e\right)^T m_\nu \right) 
+ r_2  \left(Y_e^\dagger Y_e\right)^T m_\nu \left(Y_e^\dagger Y_e\right) \nn \\
&+& r_3 \tr[Y_e^\dagger Y_e] \left( 
m_\nu \left(Y_e^\dagger Y_e\right) + \left(Y_e^\dagger Y_e\right)^T m_\nu \right)
+ r_4  \tr[Y_e^\dagger Y_e Y_e^\dagger Y_e] m_\nu \nn \\
&+& r_5 \left( m_\nu \left(Y_e^\dagger Y_e\right) 
+ \left(Y_e^\dagger Y_e\right)^T m_\nu \right) + r_6 m_\nu \; ,
\eeqa
where $r_1, r_2, r_3, r_4$ are expected to be of ${\cal O}(1)$, while 
the general forms of $r_5, r_6$ are
\beqa
r_5 &=& r_5^U \tr[Y_U^\dagger Y_U] +r_5^D \tr[Y_D^\dagger Y_D] + r_5^{g_1} g_1^2 +  
r_5^{g_2} g_2^2 + r_5^\lambda \lambda \; , \\
r_6 &=& r_{6}^{UU} \tr[Y_U^\dagger Y_U Y_U^\dagger Y_U]
+ r_{6}^{DD} \tr[Y_D^\dagger Y_D Y_D^\dagger Y_D] 
+ r_{6}^{UD} \tr[Y_U^\dagger Y_U Y_D^\dagger Y_D]  \nn \\
&+& \left( r_6^{g_1 U} g_1^2 + r_6^{g_2 U} g_2^2 + r_6^{g_3 U} g_3^2 \right) \tr[Y_U^\dagger Y_U]
+  \left( r_6^{g_1 D} g_1^2 + r_6^{g_2 D} g_2^2 + r_6^{g_3 D} g_3^2 \right) \tr[Y_D^\dagger Y_D] \nn \\
&+& \left( r_6^{g_1 \lambda} g_1^2 + r_6^{g_2 \lambda} g_2^2 \right) \lambda 
+ r_6^{\lambda} \lambda^2 + r_6^{g_1} g_1^4 + r_6^{g_2} g_2^4 + r_6^{g_1 g_2} g_1^2 g_2^2  \; ,
\label{r6}
\eeqa
with all $r_5^x, r_6^x$ being expected to be ${\cal O}(1)$ numbers.
We can further simplify by considering the fact that terms 
proportional to $\tr[Y_i^\dagger Y_i] (i \in \{ e,U, D \})$ come 
through a complete fermion loop in Higgs self-energy corrections and 
hence we must have
\beq
r_3 \tr[Y_e^\dagger Y_e] + r_5^U \tr[Y_U^\dagger Y_U] +r_5^D \tr[Y_D^\dagger Y_D] \to r_5^T \, T^\prime \; , \nn
\eeq
where $T^\prime$ is defined in Eq.~(\ref{Tprime}) and $r_5^T$ is of 
${\cal O}(1)$. So the 2-loop contribution to $\dot m_\nu$ becomes
 \beqa
\left. (4\pi)^2 \;\dot m_\nu \right\arrowvert_{\text{2-loop}} &=& 
r_1 \left( m_\nu \left(Y_e^\dagger Y_e Y_e^\dagger Y_e\right) + 
\left(Y_e^\dagger Y_e Y_e^\dagger Y_e\right)^T m_\nu \right) 
+ r_2  \left(Y_e^\dagger Y_e\right)^T m_\nu \left(Y_e^\dagger Y_e\right) \nn \\
&+& r_4  \tr[Y_e^\dagger Y_e Y_e^\dagger Y_e] m_\nu 
+ r_5^\prime \left( m_\nu \left(Y_e^\dagger Y_e\right) 
+ \left(Y_e^\dagger Y_e\right)^T m_\nu \right) + r_6 m_\nu \; ,
\label{mnu-2loop}
\eeqa
with
\beqa
r_5^\prime &=& r_5^T \, T^\prime + r_5^{g_1} g_1^2 + r_5^{g_2} g_2^2 + r_5^\lambda \lambda \; . 
\eeqa

\subsection{Results}

First, let us consider the case of the SM. Second order contributions to the
RG evolution equations of $Y_e$, $Y_\nu$, $M_\nu$ or $m_\nu$ are not
available in the literature for right-handed neutrino extended SM
(Type-I or Type-III) in general.
In Ref.~\cite{Isidori-2loop}, the contribution 
to $\dot m_\nu$ proportional to $r_2$ in Eq.~(\ref{mnu-2loop}), 
for Type-I seesaw, is presented that gives
\beq
r_2 = 2 \; .
\eeq
Thus $r_2$ is of ${\cal O}(1)$, as expected. In the future, 
once a full calculation is done, it can be checked against our results.

Next, we move to the case of the MSSM. Unlike the case of SM, there are
existing results for second order contributions in extended MSSM for
Type-I seesaw
\cite{MSSM-2nd-order}, obtained from exact computations.
In order to compare the results with the equations 
obtained above, we keep the following facts in mind:
\begin{itemize}

\item Higgs self-coupling $\lambda$ is absent in MSSM, hence 
all terms proportional to $\lambda$ will vanish.

\item Terms with $Y_e^\dagger Y_e$ can only contain 
$\tr[Y_e^\dagger Y_e]$ and $\tr[Y_D^\dagger Y_D]$, while terms with 
$Y_\nu^\dagger Y_\nu$ can only contain $\tr[Y_\nu^\dagger Y_\nu]$ and 
$\tr[Y_U^\dagger Y_U]$. Moreover they will appear only 
in the combinations $T_U$ and $T_D$, defined in 
Eqs.(\ref{TUMSSM}) and (\ref{TDMSSM}) respectively.
Thus, the terms present will be $T_D Y_e^\dagger Y_e$ and 
$T_U Y_\nu^\dagger Y_\nu$.

\item $\dot Y_e$ cannot have terms proportional to $\tr[Y_\nu^\dagger Y_\nu Y_\nu^\dagger Y_\nu]$ 
or $\tr[Y_U^\dagger Y_U Y_U^\dagger Y_U]$. Similarly, $\dot Y_\nu$
cannot have terms proportional to $\tr[Y_e^\dagger Y_e Y_e^\dagger
Y_e]$ and $\tr[Y_D^\dagger Y_D Y_D^\dagger Y_D]$. Hence
\beqa
d_{10} &=& 0 \; , \qquad d_{16}^{UU} = 0 \; , \qquad f_{9} = 0 \; ,  \qquad f_{16}^{DD} = 0 \; .
\eeqa

\item Since $Y_e$ couples to $H_D$ only, $\dot Y_e$ cannot contain terms 
$g_i^2 \tr[Y_\nu^\dagger Y_\nu]$ and $g_i^2 \tr[Y_U^\dagger Y_U]$. 
Similarly, $\dot Y_\nu$ cannot have terms $g_i^2 \tr[Y_e^\dagger Y_e]$ 
or $g_i^2 \tr[Y_D^\dagger Y_D]$.
Thus,
\beqa
d_{15}^{g_1} &=& d_{15}^{g_2} = 0 \; , \qquad  d_{16}^{g_1 U} = d_{16}^{g_2 U} = d_{16}^{g_3 U} = 0 \; \nn \\
f_{14}^{g_1} &=& f_{14}^{g_2} = 0 \; , \qquad  f_{16}^{g_1 D} = f_{16}^{g_2 D} = f_{16}^{g_3 D} = 0 \; .
\eeqa

\item The right-handed Majorana mass $M_\nu$ couples only to 
the right-handed neutrinos which interacts with $H_U$, and not with $H_D$, 
and so $h_4$ in Eq.~(\ref{h4}) becomes
\beq
h_4 = h_4^{g_1} g_1^2 + h_4^{g_2} g_2^2 + h_4^T T_U \; .
\eeq

\item Only $H_U$ is involved in the definition of the effective 
left-handed Majorana mass $m_\nu$, 
and hence we must have
\beq
T^\prime = T_U^\prime = \tr[Y_e^\dagger Y_e] + 3 \tr[Y_U^\dagger Y_U] \; ,
\eeq 
and $r_6$ in Eq.~(\ref{r6}) will have
\beqa
r_{6}^{DD} = 0 \; , \qquad  r_6^{g_1 D} = 0 \; , \qquad r_6^{g_2 D} = 0 \; , \qquad r_6^{g_3 D} = 0 \; .
\eeqa

\end{itemize}

Let us now compare the coefficients with the values obtained with exact computation 
\cite{MSSM-2nd-order}. For $Y_e$ evolution we get
\beqa
d_1 &=& -4 \; , \qquad d_2 = 0 \; , \qquad d_3 = -2 \; , \qquad d_4 = -2 \; , \qquad 
d_9 = -3 \; , \qquad d_{11} = -1 \;,\nn \\
d_{12}^{g_1} &=& 0 \; , \qquad d_{12}^{g_2} = 6 \; , \qquad d_{12}^T = -3 \; , \qquad 
d_{13}^{g_1} = 0 \; , \qquad d_{13}^{g_2} = 0 \; , \qquad d_{13}^T = -1\;, \nn \\
d_{14}^{g_1} &=& \frac{6}{5} \; , \qquad d_{14}^{g_2} = 0 \; , \qquad
d_{16}^{DD} = -9 \; , \qquad d_{16}^{UD} = -3 \; , \qquad d_{16}^{g_1 D} = -\frac{2}{5} \; , \nn \\
d_{16}^{g_2 D} &=& 0 \; , \qquad d_{16}^{g_3 D} = 16 \; , \qquad d_{16}^{g_1} = \frac{27}{2} \; , 
\qquad d_{16}^{g_2} = \frac{15}{2} \; , \qquad d_{16}^{g_1 g_2} = \frac{9}{5} \; . 
\label{di-mssm}
\eeqa

Comparing the coefficients of $\dot Y_\nu$, we get
\beqa
f_1 &=& -2 \; , \qquad f_2 = -2 \; , \qquad f_3 = 0 \; , \qquad f_4 = -4 \; , \qquad
f_{10} = -3 \; , \qquad f_{11} = -1 \; , \nn \\
f_{12}^{g_1} &=& \frac{6}{5} \; , \qquad f_{12}^{g_2} = 0 \; , \qquad f_{12}^T = -1 \; , \qquad
f_{13}^{g_1} = \frac{6}{5} \; , \qquad f_{13}^{g_2} = 6 \; , \qquad f_{13}^T = -3 \; , \nn \\
f_{15}^{g_1} &=& 0 \; , \qquad f_{15}^{g_2} = 0 \; , \qquad f_{16}^{UU} = -9 \; , \qquad 
f_{16}^{UD} = -3 \; , \qquad f_{16}^{g_1 U} = \frac{4}{5} \; , \nn \\
f_{16}^{g_2 U} &=& 0 \; ,\qquad f_{16}^{g_3 U} = 16 \; , \qquad f_{16}^{g_1} = \frac{207}{50} 
\; , \qquad f_{16}^{g_2} =  \frac{15}{2} \; , \qquad f_{16}^{g_1 g_2} =  \frac{9}{5} \; .
\label{fi-mssm}
\eeqa

Finally, comparing the evolution of $M_\nu$ and $m_\nu$ at second order, 
we find the values of $h_i$ and $r_i$s to be
\beqa
h_1 &=& -2 \; , \qquad h_2 = -2 \; , \qquad h_3 = 0 \; , \qquad 
h_4^{g_1} = \frac{6}{5} \; , \qquad h_4^{g_2} = 6 \; , \qquad h_4^T = -2 \; , 
\nn \\
r_1 &=& -2 \; , \qquad r_2 = 0 \; , \qquad r_4 = 0 \; , \qquad
r_5^{g_1} = \frac{6}{5} \; , \qquad r_5^{g_2} = 0 \; , \qquad r_5^T = -1 \; , \nn \\
r_6^{UU} &=& -18 \; , \qquad r_6^{UD} = -6 \; , \qquad r_6^{g_1 U} = \frac{8}{5} \; , \qquad r_6^{g_2 U} = 0 
\; , \qquad r_6^{g_3 U} = 32 \;,\nn \\
r_6^{g_1} &=& \frac{207}{25} \; , \qquad r_6^{g_2} = 15 \; , \qquad r_6^{g_1 g_2} = \frac{18}{5} \; .
\label{ri-mssm}
\eeqa
As we can see from Eqs.~(\ref{di-mssm}), (\ref{fi-mssm}), and
(\ref{ri-mssm}), there are a few zeros. If supersymmetry is 
not broken, one has $r_2 = 0$ in MSSM Type-I seesaw~\cite{Isidori-2loop}. 
However, the remaining zeros cannot be explained 
using spurion techniques. 
There are also some quantities which are not of
${\cal O}(1)$, namely $d_{12}^{g_2}, d_{16}^{DD},
d_{16}^{g_3 D}, d_{16}^{g_1}, d_{16}^{g_2}$, $f_{13}^{g_2},
f_{16}^{UU}, f_{16}^{g_3 U}, f_{16}^{g_2}$, $r_6^{UU}, r_6^{UD},
r_6^{g_3 U}, r_6^{g_1}$ and $r_6^{g_2}$. Of them, $x_i^{UU},
x_i^{UD}, x_i^{DD}, x_i^{g_3 U}, x_i^{g_3 D}$ can be large due to
color factors, while the remaining become large because of the effect
of gauge interactions.
%%%%%%%%%%%%%%%%%%%%%%%%%%%%%%%%%%%%%%%%%%%%%%%%%%%%%%%%
%%%%%%%%%%%%%%%%%%%%%%%%%%%%%%%%%%%%%%%%%%%%%%%%%%%%%%%%

\end{document}